\documentclass[letterpaper]{article} % DO NOT CHANGE THIS
\usepackage{sty/aaai2026}
\usepackage{times}  % DO NOT CHANGE THIS
\usepackage{helvet}  % DO NOT CHANGE THIS
\usepackage{courier}  % DO NOT CHANGE THIS
\usepackage[hyphens]{url}  % DO NOT CHANGE THIS
\usepackage{graphicx} % DO NOT CHANGE THIS
\usepackage{natbib}  % DO NOT CHANGE THIS AND DO NOT ADD ANY OPTIONS TO IT
\usepackage{caption} % DO NOT CHANGE THIS AND DO NOT ADD ANY OPTIONS TO IT

\usepackage{multirow}

\usepackage[T1]{fontenc}
\usepackage{ifthen}
\usepackage{algorithm}
\usepackage[noEnd=true,indLines=true,commentColor=black]{sty/algpseudocodex}
\usepackage{amsmath,amssymb,amsthm}
\usepackage{booktabs}
\usepackage{cleveref}
\usepackage{siunitx}
\usetikzlibrary{calc,math}
\usepackage{sty/custom}
\usepackage{enumitem}

\nocopyright

\usepackage{newfloat}
\usepackage{listings}
\DeclareCaptionStyle{ruled}{labelfont=normalfont,labelsep=colon,strut=off} % DO NOT CHANGE THIS
\floatstyle{ruled}
\newfloat{listing}{tb}{lst}{}
\floatname{listing}{Listing}
\title{Lifelong LaCAM with Local Guidance for Lifelong MAPF}

\author{
Tomoki Arita$^{1,2}$, Keisuke Okumura$^{2}$
}
\affiliations{
$^1$Keio University, Japan\\
$^2$National Institute of Advanced Industrial Science and Technology (AIST), Japan
\\
\{arita.tomoki, okumura.k\}@aist.go.jp
}

\begin{document}

\maketitle

\begin{abstract}
\emph{Local guidance} has recently proven to be a powerful driver of empirical performance in real-time,
suboptimal multi-agent pathfinding (MAPF), improving the scalable configuration-based solver \emph{LaCAM}.
By injecting informative spatiotemporal cues around each agent,
local guidance mitigates congestion, reduces waiting, and remains scalable enough even with tight time budgets,
yielding state-of-the-art performance for \emph{one-shot} MAPF.
This study asks whether the same benefits can be lifted to the \emph{lifelong} setting (\emph{LMAPF}), where tasks arrive continuously and improvements in per-step plans can increase task completion throughput over long horizons.
We propose \emph{LLLG}, \emph{a Lifelong version of LaCAM enhanced with Local Guidance}, which employs a receding-horizon windowed planning framework and warm-starts guidance from the previous solution at each timestep.
Our method scales effectively, maintains high throughput even in compact, dense environments, and surpasses existing planners, thereby pushing the frontier of real-time, lifelong MAPF.
\end{abstract}
\section{Introduction}
A central ambition in \emph{multi-agent pathfinding (MAPF)} research is to realize capable multi-robot coordination, which in practice requires \emph{lifelong} operation known as \emph{LMAPF}~\cite{Ma2017LifelongMP, chan2024league}.
In this setting, agents do not stop after executing a single plan; rather, they continue to receive new goals (i.e., tasks) once they finish the current ones.
Then, the objective is not only to maintain collision-free motion, but also to sustain high \emph{throughput} over long horizons.
Besides, deploying algorithms in the real world demands \emph{real-time} planning: at each timestep, the solver must compute the next action in less wall-clock time than the robots need to execute that timestep. 

Practical MAPF often unfolds in compact, dense environments with tens to hundreds of agents.
In such settings, the planner frequently needs to resolve potential collisions around congested spaces within each planning window.
A leading framework, \emph{RHCR (Rolling Horizon Collision Resolution)}~\cite{li2021lifelong}, is designed to produce near-optimal plans,
yet its effectiveness remains in sparse environments. Indeed, RHCR can degrade sharply in dense settings. 
This is because it typically assumes CBS/PBS-style solvers \cite{SHARON201540, Ma2018SearchingWC} and conflicts are handled through high-level branching and constrained replanning within each planning window.
When conflicts are frequent, these repeated branch-and-replan steps can accumulate, causing severe slowdowns or failure.

In parallel, over the past several years, fast, suboptimal solvers have emerged that reliably maintain feasibility even under high density.
Notable examples include \emph{PIBT (Priority Inheritance with Backtracking)}~\cite{okumura2022priority} and \emph{LaCAM (Lazy Constraints Addition Search)}~\cite{okumura2023lacam}, both of which scale well and deliver real-time decisions in scenarios where many other methods falter.
PIBT functions as a fast one-step planner that synthesizes collective collision-free motions. 
Then, to avoid local deadlock or livelock, LaCAM solves long-horizon planning by leveraging this generator and performing a systematic search over configurations.
While PIBT and LaCAM bring feasible solutions rapidly, they remain highly suboptimal in dense settings, where purely goal-directed action repeatedly induces \emph{congestion} and degrades solution quality.

{
\setlength{\tabcolsep}{2pt}
\newcommand{\methodposDenLaCAM}{{0.17,0.14}}
\newcommand{\methodposDenLG}{{0.37,0.87}}
\newcommand{\methodposDenWPPL}{{0.24,0.22}}
\newcommand{\methodposDenRHCR}{{0.4,0.59}}
\newcommand{\methodposDenGuidedPIBT}{{0.12,0.73}}

\newcommand{\methodposEmptyLaCAM}{{0.18,0.38}}
\newcommand{\methodposEmptyLG}{{0.3,0.86}}
\newcommand{\methodposEmptyWPPL}{{0.25,0.58}}
\newcommand{\methodposEmptyRHCR}{{0.3,0.21}}
\newcommand{\methodposEmptyGuidedPIBT}{{0.3,0.48}}

\newcommand{\methodposRandomLaCAM}{{0.24,0.72}}
\newcommand{\methodposRandomLG}{{0.3,0.88}}
\newcommand{\methodposRandomWPPL}{{0.7,0.55}}
\newcommand{\methodposRandomRHCR}{{0.3,0.27}}
\newcommand{\methodposRandomGuidedPIBT}{{0.3,0.63}}

\newcommand{\methodposWarehouseLaCAM}{{0.26,0.26}}
\newcommand{\methodposWarehouseLG}{{0.27,0.86}}
\newcommand{\methodposWarehouseWPPL}{{0.27,0.70}}
\newcommand{\methodposWarehouseRHCR}{{0.27,0.65}}
\newcommand{\methodposWarehouseGuidedPIBT}{{0.27,0.25}}

\newcommand{\methodlabel}[5]{% #1: panel key, #2: method key, #3: color, #4: label, #5: anchor
  \node[anchor=#5] at (\csname methodpos#1#2\endcsname) {\small \textcolor[HTML]{#3}{#4}};
}
\newcommand{\methodlabels}[1]{% #1: panel key
  % \methodlabel{#1}{LaCAM}{008000}{LaCAM}{west}
  % \methodlabel{#1}{LG}{0000FF}{LLLG}{west}
  % \methodlabel{#1}{WPPL}{FF8C00}{WPPL}{west}
  % \methodlabel{#1}{RHCR}{800080}{RHCR}{west}
  % \methodlabel{#1}{GuidedPIBT}{ff66c4}{TFO}{west}
}
\newcommand{\xlabellegendsep}{0.2em}
\newcommand{\highlightpanelheight}{\linewidth}
\newcommand{\highlightpanelwidth}{0.24\linewidth}
\newcommand{\entry}[3][\highlightpanelwidth]{% #1: panel width, #2: instance key, #3: panel key
  \begin{minipage}[t]{#1}
    \centering
    \begin{tikzpicture}
      \node[anchor=south west,inner sep=0] (img) at (0, 0) {%
        \includegraphics[width=\linewidth]{fig/raw/highlight/runtime_throughput_#2_agents800}%
      };
      \begin{scope}[x={(img.south east)},y={(img.north west)}]
        \methodlabels{#3}
      \end{scope}
    \end{tikzpicture}
  \end{minipage}%
}
\newcommand{\ylabelbox}{\makebox[0pt][r]{\ylabel\hspace{0.4em}}}
\newcommand{\ylabel}{\rotatebox{90}{\small \hspace{0.3em}{ \hspace{0.7em} throughput $\rightarrow$}}}
\newcommand{\xlabel}{\small  $\leftarrow$ runtime [\SI{}{\second}]}
\newcommand{\maphead}[4][0pt]{%
  \hspace*{#1}%
  \begin{tabular}[c]{@{}c@{}}
    {\scriptsize \mapname{#2}}\\[-0.15em]
    {\scriptsize (#3)}
  \end{tabular}%
}
\begin{figure}[t!]
  \centering
  \begin{tabular}{@{}cccc@{}}
    \maphead[8pt]{empty}{48x48}{2,304} & \maphead[8pt]{random}{64x64}{3,684} & \maphead[4pt]{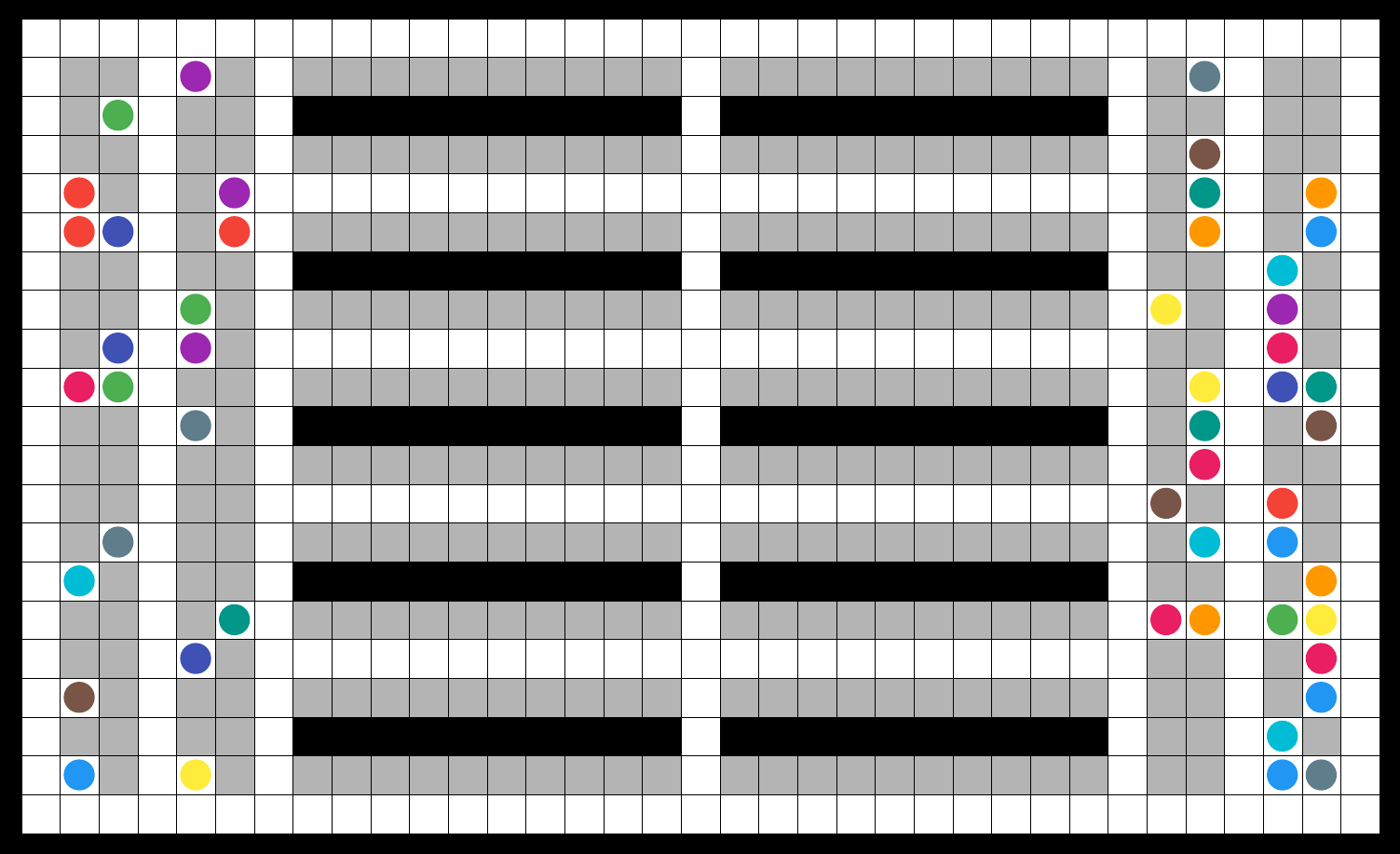}{170x84}{9,776} & \maphead[2pt]{ht_chantry}{162x141}{7,461} \\
    \ylabelbox
    \entry[0.225\linewidth]{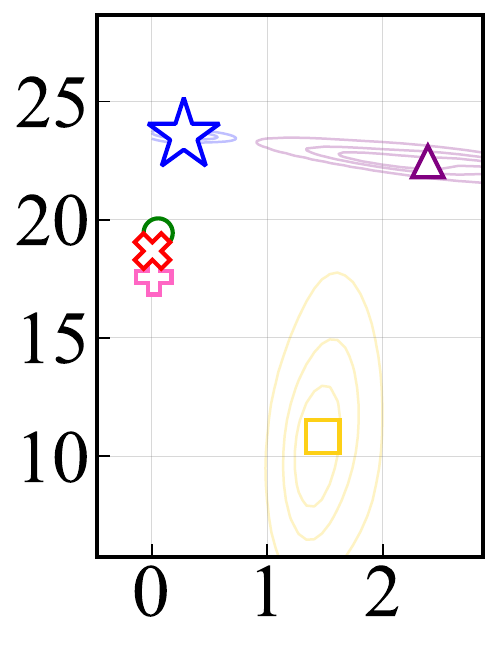}{Empty} &
    \entry[0.225\linewidth]{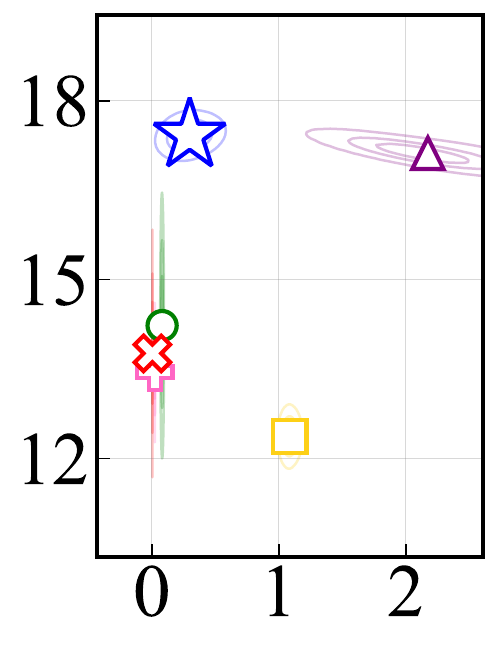}{Random} &
    \entry[0.225\linewidth]{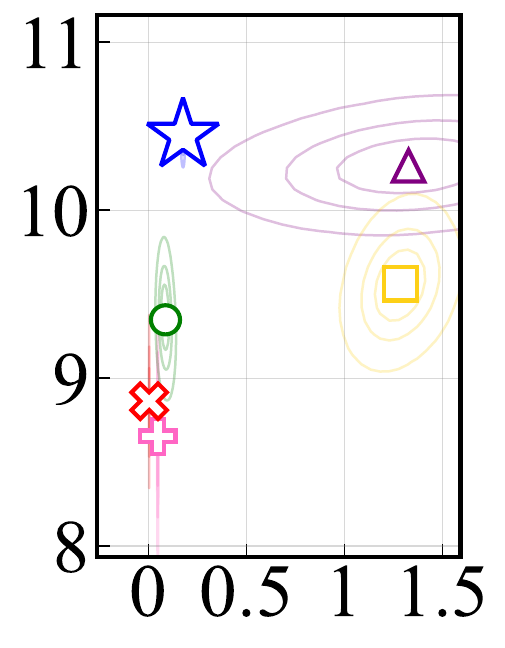}{Warehouse} &
    \entry[0.22\linewidth]{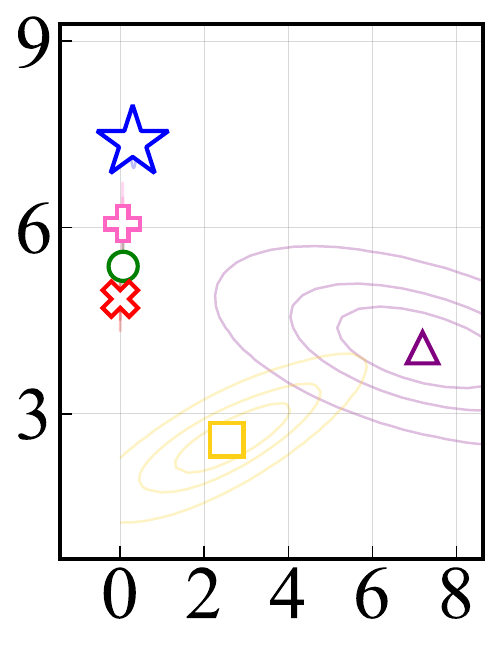}{Den}
    \\
    \multicolumn{4}{c}{\xlabel} \\
    \noalign{\vskip\xlabellegendsep}
    \multicolumn{4}{c}{\includegraphics[width=0.8\linewidth,clip,trim={0 0.8cm 0 0.5cm}]{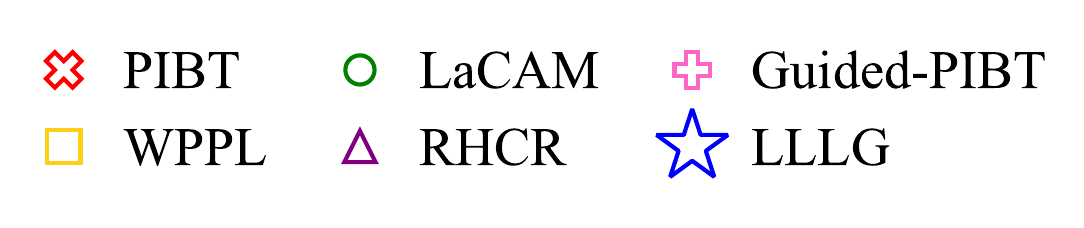}} \\
  \end{tabular}
  \caption{
    Performance of lifelong MAPF methods in representative scenarios, evaluated by throughput and runtime.
    All methods are evaluated over 500 steps with 800 agents across all maps.
  }
  \label{fig:highlight}
\end{figure}

}

In summary, RHCR produces efficient solutions, but struggles in dense environments.
In contrast, PIBT and LaCAM are stable, but highly suboptimal.
To this end, this work proposes \emph{Lifelong LaCAM with Local Guidance (LLLG)}, which quickly produces high-throughput solutions even in dense settings.
\Cref{fig:highlight} highlights results across several compact, high-density LMAPF scenarios, demonstrating that LLLG carves out a new performance frontier for LMAPF. 
For example, on \mapname{empty-48-48}, \mapname{random-64-64-10}, and \mapname{warehouse-10-20-10-2-2}, LLLG consistently attains higher throughput than RHCR while reducing runtime by approximately 90\%, with only a modest runtime overhead relative to PIBT and LaCAM.
Furthermore, on instances of \mapname{ht_chantry} with bottlenecks, where RHCR degrades sharply, LLLG achieves 81\% higher throughput while reducing runtime by 96\%.

This superior performance of LLLG is driven by \emph{guidance}, which biases the underlying solver with costs designed to mitigate emerging congestion---for example, penalizing moves into predicted high-traffic cells/edges and excessive waiting, while rewarding detours or time shifts that distribute agents across space and time.
Whereas many prior methods rely on \emph{global} guidance that is derived from a map-wide view of the environment and typically time-independent \cite{chen2024traffic, zhang2024guidance, Kato2025CongestionMP}, recent work suggests that \emph{local} guidance can be more effective by providing agent-centric spatiotemporal cues for imminent conflicts for \emph{one-shot MAPF}~\cite{arita2025local}.
Following this insight, our LMAPF planner embeds local guidance into a windowed planning, receding horizon framework.
In what follows, we present problem formulation, technical background, the proposed method, experimental evaluation, and discussion in order.
The code is publicly available at \url{https://github.com/allegorywrite/lllg}.

\section{Preliminaries}

This section defines LMAPF, reviews key planning concepts, and summarizes the technical basis of our method.

\subsection{Problem Definition}
We study lifelong multi-agent pathfinding (LMAPF) on an undirected graph $G=(V, E)$ with a team of agents $A=\{1,\ldots,n\}$.
Time is discrete, and agents move synchronously.
At each timestep, each agent either waits or moves to an adjacent vertex.
Collisions are interpreted in the standard MAPF sense~\cite{stern2019def}: two agents may not occupy the same vertex at the same timestep, nor traverse the same edge in opposite directions simultaneously.

A \emph{configuration} is the tuple of all agents' locations $\Q \in V^n$.
Let $\Q^0=(s_1,\ldots,s_n)$ be the initial configuration, and let $g_i^t \in V$ denote agent $i$'s current goal at time $t$.
After initialization, each agent is assigned a new goal whenever it reaches its current goal.
In our experiments, the goal is assigned randomly from all the vertices.
A \emph{solution} is an infinite collision-free sequence of configurations in which each agent moves to an adjacent vertex or waits at every timestep.
We evaluate solution quality by \emph{throughput}, measured as the average number of completed tasks per timestep.

\subsection{Planning Concepts for LMAPF}

LMAPF admits several design choices in how planning and execution are interleaved.
We summarize common concepts below (e.g., windowed planning) and use this taxonomy to position our method.

\paragraph{Naive Approach.}
The most straightforward strategy to LMAPF is to solve a one-shot MAPF problem at every timestep from the current configuration $\Q^t \in V^n$ to a goal configuration $\mathcal{G} = (g_1^t,\ldots,g_n^t) \in V^n$, yielding a solution $\Pi^t$, and then execute only its first joint action $\Pi^t[1]$ to obtain $\Q^{t+1}$, e.g.,~\cite{Shankar2025LFOM}.
This keeps the policy responsive to congestion and online goal updates, but it can be computationally demanding under strict real-time budgets.

\paragraph{Windowed Planning.}
To amortize computation, \emph{windowed planning} solves a finite-horizon MAPF problem over a window of length $w_{\Pi}\in \mathbb{N}_{\geq 1}$ and optimizes only behavior within this window \cite{Veerapaneni2024WindowedMW}.
Classic online solvers such as WHCA$^*$~\cite{silver2005cooperative, 6907401} reserve paths only for the next $w_{\Pi}$ timesteps, and modern lifelong methods such as WPPL~\cite{Jiang2024ScalingLM} likewise build coordination around finite-window optimization and refinement.
One extreme is $w_{\Pi}{=}1$, which reduces planning to a purely myopic, one-step decision rule (e.g., PIBT;~\citeauthor{okumura2022priority}~\citeyear{okumura2022priority}).
At the other extreme, the naive approach described above corresponds to \emph{full-horizon} planning with $w_{\Pi}{=}\infty$, whereas in lifelong settings the planning window typically spans only up to the agents' currently assigned goals.

\paragraph{Receding Horizon.}
The above discussion has focused on how to compute a plan.
For real-time lifelong operation, however, such planning can be coupled with execution in a \emph{receding horizon} manner, like model predictive control (MPC).
With this fashion, at each timestep, the solver computes a plan from the current configuration, executes the first $h\in\mathbb{N}_{\geq 1}$ steps of that plan, and then replans from the updated configuration.
Depending on the underlying planner, the computed plan may target the currently assigned goals over the full remaining horizon or only over a finite planning window (e.g., RHCR;~\citeauthor{li2021lifelong}~\citeyear{li2021lifelong}).
One appealing choice is $h{=}1$, which replans every timestep and is often reasonable in LMAPF because it is highly responsive and does not require goal lookahead.

\paragraph{Incremental Planning.}
In receding horizon planning, much of the intermediate information computed in one replanning cycle is discarded before the next begins.
Since consecutive cycles in LMAPF are often highly correlated, especially when $h{=}1$, it is beneficial to carry information across cycles rather than restart from scratch.
This idea echoes incremental heuristic search, where prior search effort is reused across a sequence of related problems such as D$^*$ Lite~\cite{Koenig2002Dlite}, and also recent one-shot MAPF frameworks such as RT-LaCAM~\cite{Liang2025RealTimeLF}.
However, unlike the one-shot setting, LMAPF frequently updates agents' goals online, making the previously constructed search tree difficult to exploit effectively.
In receding horizon settings, this same intuition naturally appears as warm starting, i.e., initializing the current planning solution with the path computed in the previous cycle, as like standard implementation of MPC.

\paragraph{Anytime Planning.}
In LMAPF, each executed action is derived from a one-shot (potentially windowed) solution and can thus benefit from \emph{anytime} planning, where an initial feasible solution is refined as more computation time becomes available.
In one-shot MAPF, such a strategy is common, as fast suboptimal solvers are well established~\cite{li2022mapf}.
Since a lifelong solution consists of a sequence of such one-shot solutions, improving each windowed plan can be expected to increase lifelong throughput~\cite{Jiang2024ScalingLM}.

\paragraph{Our Approach.}
Our method follows a receding horizon scheme with $h{=}1$, replanning at every timestep while executing only the first step of the current plan.
At each timestep, we warm start the local guidance by reusing the planning result at the previous timestep.
This produces an initial, feasible, and high-quality windowed plan quickly and also allows further anytime refinement.
To compute each one-shot plan over a horizon of length $w_{\Pi}$, we utilize LaCAM with local guidance, which we describe next.

\subsection{LaCAM and PIBT}

A natural way to solve \emph{one-shot} MAPF is to search over \emph{joint configurations}, where each node represents the locations of all agents and each edge corresponds to a one-step synchronous transition.
Namely, each search node $N$ is associated with a joint configuration $N.\mathcal{Q}\in V^n$.
However, a vanilla search in this space can suffer from a prohibitive branching factor, which can be as large as $O(5^n)$ in a grid world, since enumerating all joint successors requires considering combinations of individual agents' moves.

\emph{LaCAM}~\cite{okumura2023lacam} is a configuration-based solver that addresses this difficulty by combining a high-level search over joint configurations with \emph{lazy successor generation}.
Given $N.\mathcal{Q}$, a fast configuration generator proposes a promising collision-free next configuration $\mathcal{Q}'$ (i.e., one move for each agent) on demand, yielding a successor node $N'$ with $N'.\mathcal{Q}=\mathcal{Q}'$.

A standard choice of configuration generator in LaCAM is \emph{PIBT}~\cite{okumura2022priority}.
PIBT determines each agent's next move by ranking candidate vertices
$v \in \neigh(\Q[i]) \cup \{\Q[i]\}$, including wait.
The ranking is based on the lexicographic cost
$\langle \dist(v,g_i), \epsilon \rangle$,
with smaller costs preferred, and PIBT selects the first collision-free option in that order.
Here, $\dist(v,g_i)$ is the shortest-path distance from $v$ to the current goal $g_i$, while $\epsilon$ is a random tiebreak term.

\subsection{LG-LaCAM}

In high-density environments, purely goal-directed local decisions based on $\dist(v,g_i)$ can concentrate many agents onto moves that appear individually favorable, thereby creating \emph{congestion} even when each local choice is reasonable.
This, in turn, increases the need for collision resolution, degrading solution quality.
\emph{LG-LaCAM}~\cite{arita2025local} addresses this issue by augmenting configuration generation with \emph{local guidance}: for each planning step, it constructs agent-wise guidance paths $\Phi$ over a $w_{\Phi}$-step window so as to mitigate anticipated local congestion.
Intuitively, the objective is to find, for each agent, a short path fragment that minimizes collisions within the window without becoming overly conservative.
More precisely, with a hyperparameter $\alpha \in \mathbb{R}_{\geq 0}$, a single-agent guidance path for agent $i\in A$  is constructed by solving
\begin{equation}
\begin{aligned}
\argmin_{\pi \in \Omega} \: \cost_i(\pi)
&= \langle\dist(\pi[w_{\Phi}], g_i), 0\rangle \\
&\quad + \sum_{t = 0}^{w_{\Phi}-1}
\langle 1 + \alpha \cdot \funcname{Ind}[\chi > 0], \chi \rangle .
\end{aligned}
\label{eq:guidance}
\end{equation}
where $\Omega$ denotes the set of paths of length $w_{\Phi}{+}1$ starting from the current location $Q[i]$, and $\chi$ is the number of collisions of the transition $(\pi[t],\pi[t+1])$ with other paths currently stored in $\Phi$.
Here, \funcname{Ind} is an indicator function that returns one only if the condition is true, and zero otherwise.
These guidance paths can be computed by space-time A$^\ast$.

\Cref{algo:guidance} outlines how to construct the guidance paths $\Phi$, which is initialized as empty at the initial timestep, whereas at later timesteps it is warm-started from the guidance of the parent node.
Then, the guidance is refined sequentially for each agent, and this procedure is repeated $m$ times to reduce bias induced by the agent update order.
The resulting guidance is then injected into PIBT through the lexicographic cost
$\langle \funcname{Ind}[\Phi[i][1] \neq v], \dist(v,g_i), \epsilon \rangle$,
which first prefers moves consistent with the guidance and then preserves the original goal-directed preference.

{
\begin{algorithm}[th!]
\caption{Guidance Construction}
\label{algo:guidance}
\begin{algorithmic}[1]
\small
\Input{configuration $\Q \in V^n$, goals $\G = (g_1, g_2, \ldots, g_n) \in V^n$}
\Output{$\Phi \in V^{n\times (w_{\Phi} + 1)}$\quad\textbf{params:}~$w_{\Phi}, m \in \mathbb{N}_{>0}$}
\State initialize $\Phi$
\label{algo:guidance:init}
\For{$r = 1,2,\ldots,m$} \Comment{guidance refinement}
\label{algo:guidance:termination}
\For{$i \in A$}
\label{algo:guidance:order}
\State update $\Phi[i]$ by solving \Cref{eq:guidance}
\label{algo:guidance:pathfinding}
\EndFor
\label{algo:guidance:for-end}
\EndFor
\State \Return $\Phi$
\end{algorithmic}
\end{algorithm}
}

\subsection{LaCAM$^*$}
\emph{LaCAM$^\ast$}~\cite{okumura2023lacam2} is an anytime variant of LaCAM that improves an initial feasible plan over time and can converge to an optimal solution under cumulative transition-cost objectives.
After obtaining an initial solution, it continues searching for lower-cost ones using branch-and-bound; it maintains the current best plan and prunes nodes that cannot improve it.
Specifically, each node $N$ is scored by $f(N)=g(N)+h(N.\mathcal{Q})$, where $g(N)$ is the accumulated transition cost to $N$ and $h(N.\mathcal{Q})$ is a heuristic estimate of the remaining cost to reach the goal configuration.

\section{Method}
Our idea is to leverage local guidance from one-shot MAPF to LMAPF.
This adaptation, however, is nontrivial because goals change upon completion, preventing direct reuse of one-shot plans.
Meanwhile, because many agents retain the same goals across consecutive timesteps, much of the previous plan can be reused with a simple time shift, making such reuse crucial for performance.
To address this, we propose \emph{Lifelong LaCAM with Local Guidance (LLLG)}, 
a receding-horizon framework that replans over a finite window while \emph{warm-starting local guidance} from the previous timestep, rather than directly warm-starting the current solution.

\paragraph{Structure.}
As discussed preliminarily, a natural starting point for extending local guidance to LMAPF is the naive approach: at every timestep, solve a one-shot MAPF instance from the current configuration using LG-LaCAM and execute only the first joint action.
This choice is attractive because LG-LaCAM is a powerful one-shot solver in dense settings, but replanning the full-horizon problem from scratch at every timestep can be computationally expensive under real-time constraints.

To reduce this burden, we introduce \emph{windowed planning}.
Instead of solving the full-horizon planning at each timestep, we run LG-LaCAM with its high-level search truncated at depth $w_{\Pi}$.
When the search reaches depth $w_{\Pi}$, we stop expanding the node and regard the backtracked sequence of configurations from the root to that node as a $w_{\Pi}$-step windowed plan.
This reduces the cost of replanning while still allowing the solver to mitigate short-term congestion,
which is important in lifelong settings where goals continue to change over time, making long-term planning uncertain.
While windowed LaCAM looks similar to rolling out PIBT $w_{\Phi}$ steps, we argue that employing LaCAM search provides advanced search techniques, e.g., preventing local livelock, LaCAM$^\ast$, and deadlock detection~\cite{Jain2025GraphAS}.
We then embed this windowed solver into a \emph{receding horizon} framework.
At each timestep, the planner computes a windowed plan from the current configuration, executes only the first timestep, and then replans after one timestep is executed.

A key consideration is how to exploit information from the previous timestep, given the substantial overlap between consecutive replanning cycles, as such reuse can reduce per-step computation cost.
As illustrated in \cref{fig:concept}, in the configuration generation of LG-LaCAM, LLLG uses warm-started guidance associated with each configuration.
At the root configuration, we initialize the guidance paths $\Phi$ from the suffix of the previous timestep's solution $\Pi^{t-1}[2{:}w_{\Phi}]$.
For successor nodes during high-level search, we initialize the guidance from one computed at the parent node.
In this way,  each planning cycle leverages the information from the previous cycle, instead of completely discarding the previous planning effort.

Finally, because the resulting windowed plan can be improved sometimes within the same per-step time budget, we can optionally incorporate \emph{anytime refinement}.
After obtaining an initial feasible $w_{\Pi}$-step plan, the planner continues improving it using either LaCAM$^\ast$ or LNS.
We defer the details of these refinements to later.
These components together yield LLLG, summarized in \cref{alg:lifelong-lacam}.

{
  \begin{figure}[t!]
    \centering
    \includegraphics[width=1.0\linewidth,clip,trim={5.5cm 3.5cm 7.5cm 3.2cm}]{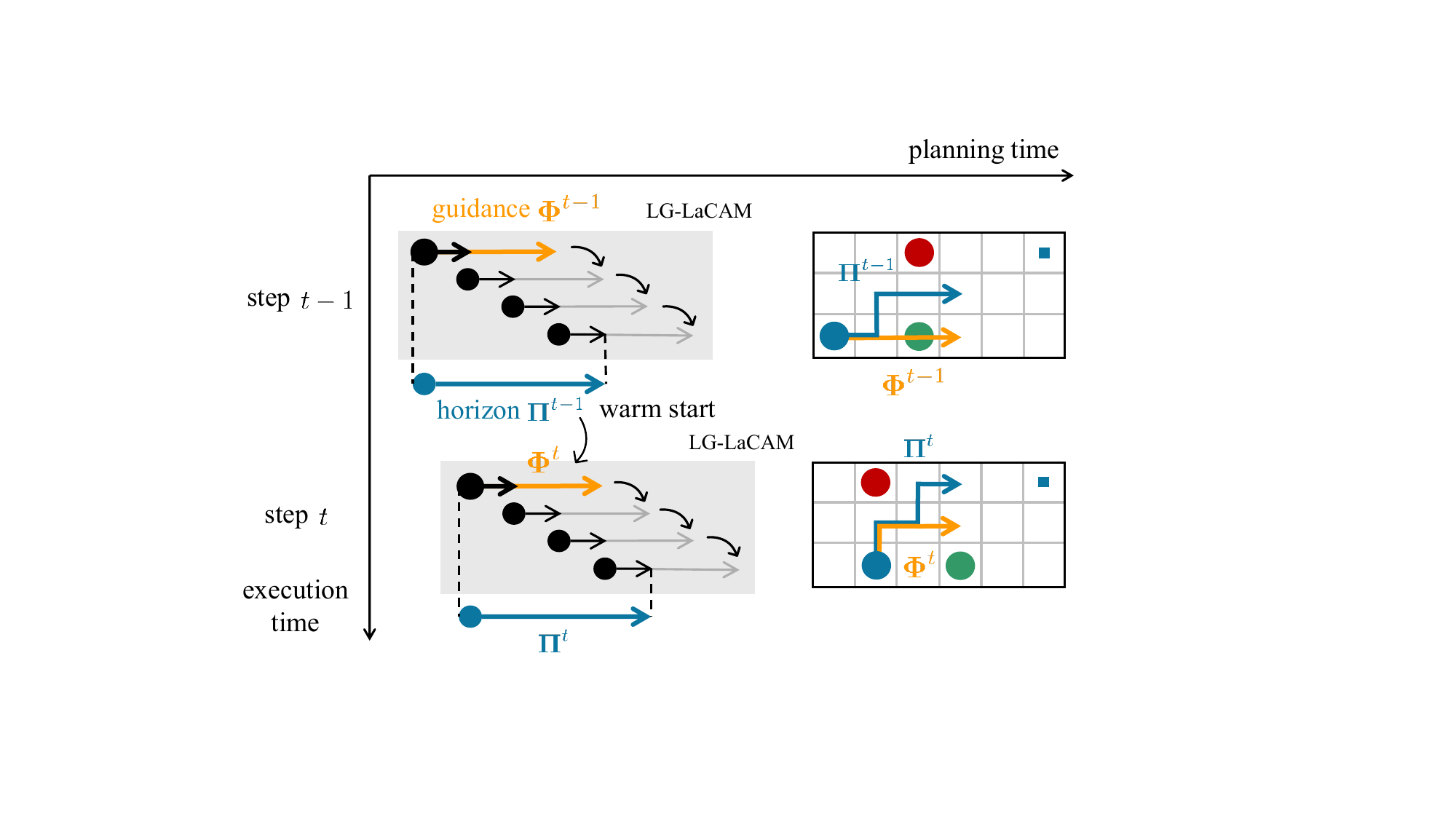}
\caption{
Concept of LLLG. 
The gray-shaded regions correspond to configuration generation by LaCAM.
At each timestep, the initial guidance $\boldsymbol{\Phi}$ is warm-started from the previous step's solution $\boldsymbol{\Pi}$.
}
    \label{fig:concept}
  \end{figure}
}

{
\begin{algorithm}
\caption{Lifelong LaCAM with Local Guidance}
\label{alg:lifelong-lacam}
\begin{algorithmic}[1]
\While{not terminated}
  \State update goals $\mathcal{G}$
  \State initialize $\Phi^t$ with previous solution $\Pi^{t-1}[2{:}w_{\Phi}]$
  \State $\Pi^t \gets$ windowed LG-LaCAM for $(\mathcal{Q}^t, \mathcal{G})$
  \State optionally refine $\Pi^t$ by \{LaCAM$^*$, LNS\}
  \State execute $\Pi^t[1]$
\EndWhile
\end{algorithmic}
\end{algorithm}
}

We now detail the key design choices of our method.

\paragraph{Leveraging LG in Receding Horizon.}
Under receding horizon execution, the guidance $\Phi$ is recomputed at every timestep, and performance depends on how we initialize each replanning cycle by reusing information from the previous step's planning result.
Two warm-start schemes of LLLG are possible:
\begin{enumerate}[label=(\roman*)]
  \item Initializing $\Phi^t$ with $\Phi^{t-1}$ at the root configuration of the previous timestep, denoted LLLG$_{\boldsymbol{\Phi}}$.
  \item Initializing the next guidance from the suffix of the previous finite-horizon solution $\Pi^{t-1}$, denoted LLLG$_{\boldsymbol{\Pi}}$.
\end{enumerate}
For (ii), the guidance is initialized from the suffix of $\Pi^{t-1}$, using $\Pi^{t-1}[2{:}w_{\Phi}]$ when $w_{\Phi}{\leq} w_{\Pi}$; otherwise, we use $\Pi^{t-1}[2{:}w_{\Pi}]$ and pad the remaining steps up to length $w_{\Phi}$ with the terminal configuration.
Later, we will evaluate these two warm-start schemes and show that (ii) is more effective in our lifelong receding horizon setting.
Therefore, unless otherwise noted, LLLG refers to LLLG$_{\boldsymbol{\Pi}}$.

\paragraph{Refinement of local guidance.}
In \cref{algo:guidance}, local guidance construction updates agent-wise paths sequentially;
the resulting guidance can be biased by the update order within a single pass.
Following LG-LaCAM~\cite{arita2025local}, we perform $m$ rounds of guidance refinement within each timestep.
This reduces order-induced bias and improves solution quality.
In the original full-horizon one-shot setting, LG-LaCAM uses $m{=}1$ because repeated refinement is costly when guidance must be refined at every step along the path from start to goal.
By contrast, LLLG replans only over a finite window, making additional refinement practical.
Accordingly, we typically use $m{=}2$ in LLLG.

\paragraph{Refinement of planning path.}
At each timestep, after LG-LaCAM returns an initial $w_{\Pi}$-step windowed plan, the planner can continue refining this plan until the next wall-clock timestep arrives.
Because the windowed plan quality affects both the executed first action and the guidance carried over to the next timestep, improvements in one-shot windowed solution quality may serve as a promising indicator for higher lifelong throughput.
We consider two refinement options, each representing a design choice for allocating computation within a receding horizon planner.

\begin{itemize}
  \item \textbf{LaCAM$^*$}~\cite{okumura2023lacam2} continues the horizon-limited LaCAM search after a first feasible plan is found.
  The planner keeps expanding its search within the same $w_{\Pi}$-step window and returns the best plan found under the time limit.
  In our implementation, the $g$-value is accumulated by a per-step transition cost
  $\cost(\mathcal{Q}^{-},\mathcal{Q})=\sum_{i\in A}\funcname{Ind}[\mathcal{Q}[i]\neq \mathcal{Q}^{-}[i]\ \lor\ \mathcal{Q}^{-}[i]\neq g_i]$,
  and the heuristic is $h(\mathcal{Q})=\sum_{i\in A}\dist(\mathcal{Q}[i],g_i)$.
  \item \textbf{Large Neighborhood Search (LNS)} \cite{li2021anytime} refines the current $w_{\Pi}$-step joint plan by repeatedly selecting a subset of agents, removing their paths, and replanning only for that subset while treating the remaining paths as fixed obstacles in space time \cite{Jiang2024ScalingLM}.
  Replanning is performed using a finite-horizon space-time A$^\ast$ search\footnote{Or SIPP~\cite{phillips2011sipp}.} over states $(v,t)$ with evaluation function $f=g+h$.
  The path cost is accumulated using the per-step cost $\cost(v,t)$, defined as $\cost(v,t)=1$ if the agent has not yet reached its goal and $\cost(v,t)=0$ otherwise.
  The heuristic is defined as $h(v,t)=\dist(v,g_i)$ before the goal is reached and $h(v,t)=0$ afterward.
\end{itemize}

\paragraph{Integration with Hindrance.}
We further refine PIBT's preference construction by integrating \emph{hindrance}~\cite{okumura2025lightweight}, a lightweight one-step estimate of how an agent's move may hinder nearby agents at the next timestep, thereby mitigating short-term blocking.
We combine hindrance with local guidance by ranking candidate moves $u \in \neigh(\Q[i]) \cup \{\Q[i]\}$ using a cost
$\langle \funcname{Ind}[\Phi[i][1] \neq u], \dist(u,g_i), \funcname{hindrance}, \epsilon \rangle$.
An empirical result of this effect is available in the appendix.

\section{Evaluation}
Experiments were conducted on a Mac Studio with M1 Ultra \SI{64}{\giga\byte} of RAM.
We use maps from the MAPF benchmark~\cite{stern2019def}, and use the provided initial configurations as the start states.
For evaluation, we report throughput and runtime, defined as the average number of completed tasks and computation time per executed timestep.
Unless specified otherwise, each method is evaluated over 500 steps and 10 instances with a per-step planning timeout of \SI{10}{\second}.
We use this setting to broaden the evaluation by examining behavior under longer per-step planning times.
We compare representative leading approaches for LMAPF as follows:

\begin{itemize}
  \item \textbf{RHCR}~\cite{li2021lifelong} is a leading LMAPF framework that interleaves execution and planning by solving windowed MAPF subproblems online. 
  In our experiments, we use \emph{Priority-Based Search (PBS)}~\cite{Ma2018SearchingWC} as the underlying MAPF solver, with the execution length set to $h{=}5$ and the planning horizon set to $w_{\Pi}{=}10$.
  \item \textbf{PIBT}~\cite{okumura2022priority} (without hindrance) is a fundamental approach in the line of suboptimal methods for MAPF and its lifelong variant, LMAPF. 
  \item \textbf{(Lifelong) LaCAM} denotes our receding horizon extension of LaCAM~\cite{okumura2023lacam} without any guidance terms, using full-horizon planning. 
  \item \textbf{Guided-PIBT}~\cite{chen2024traffic} augments PIBT with congestion-avoiding guidance to improve throughput in the lifelong setting.
  \item \textbf{WPPL}~\cite{Jiang2024ScalingLM} is a windowed anytime lifelong solver that combines PIBT with LNS, and it was the champion method in the first LMAPF competition~\cite{chan2024league}.
  We evaluate two variants, using \SI{1}{\second} and \SI{10}{\second} of LNS refinement per timestep as specified in the official implementation, both without either Manual guidance or Guidance Graph Optimization.
  \item \textbf{Our proposed method, LLLG.} Unless noted otherwise, we set $w_{\Phi}{=}20$, $w_{\Pi}{=}10$, and $m{=}2$.
\end{itemize}
For the baselines, the authors' provided codes are used.%
\footnote{
RHCR: \url{https://github.com/Jiaoyang-Li/RHCR};
PIBT: \url{https://github.com/HirokiNagai-39/pibt-tiebreaking};
Guided-PIBT: \url{https://github.com/nobodyczcz/guided-pibt};
WPPL: \url{https://github.com/DiligentPanda/MAPF-LRR2023}
}
We exclude learning-based approaches, such as \cite{jiang2025deploying}, from comparison since they constitute a different methodological category from the heuristic approaches considered in this study.
For all LaCAM-based methods, we use the initial solution returned at each timestep (i.e., without anytime refinement) and enable the hindrance tiebreak.

{
\begin{figure*}[t!]
\centering

\begin{minipage}{\linewidth}
\centering
{
\setlength{\tabcolsep}{1pt}
\newcommand{\heatmapfile}[1]{fig/raw/heatmap/heatmap_stops_random-64-64-10_result_#1_random-64-64-10_1000_random-64-64-10-random-1}
\newcommand{\histogramfile}[1]{fig/raw/heatmap/heatmap_stops_random-64-64-10_result_#1_random-64-64-10_1000_random-64-64-10-random-1_histogram}
\newcommand{\splitkeyval}[1]{\splitkeyvalaux#1\relax}
\def\splitkeyvalaux#1=#2\relax{#1 & = & #2}
\newcommand{\metricstwo}[2]{%
{\renewcommand{\arraystretch}{0.9}%
\begin{tabular}{@{}l@{\,}c@{\,}l@{}}
\splitkeyval{#1}\\
\splitkeyval{#2}
\end{tabular}}%
}
\newcommand{\entrymetrics}[2]{%
\if\relax\detokenize{#1}\relax
\else
\node[anchor=north] at (2.3, 4.5) {\small \metricstwo{#1}{#2}};
\fi
}
\newcommand{\entry}[5]{
\begin{scope}[xshift=0.19*#3\linewidth, scale=0.82, transform shape]
{
\node[anchor=south west] at (0.5, -0.1)
{\includegraphics[width=0.2\linewidth]{\heatmapfile{#2}}};
\node[] at (2.3, 4.75) {\small #1};
\entrymetrics{#4}{#5}
\node[anchor=north west] at (0.65, -0.0)
{\includegraphics[width=0.2\linewidth,clip,trim={1.6cm 1.3cm 0 0}]{\histogramfile{#2}}};

\node[anchor=north east, overlay] at (0.9, -0.65){\scriptsize 0};
\node[anchor=north east, overlay] at (0.9, -0.45){\scriptsize 1};
\node[anchor=north east, overlay] at (0.9, -0.26){\scriptsize 10};
\node[anchor=north east, overlay] at (0.9, -0.07){\scriptsize 100};
\node[anchor=north, overlay] at (0.9, -0.8){\scriptsize 1};
\node[anchor=north, overlay] at (1.45, -0.8){\scriptsize 20};
\node[anchor=north, overlay] at (2.1, -0.8){\scriptsize 40};
\node[anchor=north, overlay] at (2.8, -0.8){\scriptsize 60};
\node[anchor=north, overlay] at (3.5, -0.8){\scriptsize 80};
\node[anchor=north, overlay] at (4.10, -0.8){\scriptsize 100};

\node[anchor=north east,rotate=0] at (0.9, 0.15) {\scriptsize freq.};
\node[anchor=north west] at (1.5, -1.02) {\scriptsize number of stops};
}
\end{scope}
}
\textbf{\mapname{random-64-64-10},  1,000 agents}\par\vspace{0.2em}
\begin{tikzpicture}
\entry{RHCR}{RHCR}{0}{throughput=20.7}{runtime=2.13 [s/step]}
\entry{PIBT}{pibt_tiebreaking}{1}{throughput=14.7}{runtime=0.0015 [s/step]}
\entry{LaCAM}{lifelong_lacam}{2}{throughput=17.1}{runtime=0.043 [s/step]}
\entry{Guided-PIBT}{guided_pibt}{3}{throughput=16.0}{runtime=0.015 [s/step]}
\entry{LLLG (proposed)}{lifelong_lg_lacam}{4}{throughput=\textbf{21.1}}{runtime=0.21 [s/step]}
\node[anchor=south west] at (0.95\linewidth, -0.0)
{\includegraphics[width=0.028\linewidth]{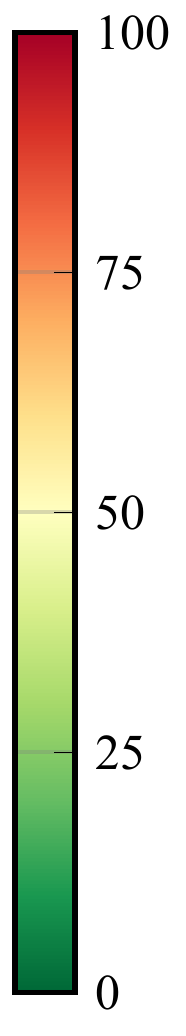}};
\node[anchor=south east,rotate=-90] at (0.985\linewidth, 0.45) {\small number of stops};
\end{tikzpicture}
}
\end{minipage}

\vspace{-0.6em}

\begin{minipage}{\linewidth}
\centering
{
\setlength{\tabcolsep}{1pt}
\newcommand{\heatmapfile}[1]{fig/raw/heatmap2/heatmap_stops_empty-48-48_result_#1_empty-48-48_1000_empty-48-48-random-1}
\newcommand{\histogramfile}[1]{fig/raw/heatmap2/heatmap_stops_empty-48-48_result_#1_empty-48-48_1000_empty-48-48-random-1_histogram}
\newcommand{\histtrimdefault}{1.5cm 1.3cm 0 0}
\newcommand{\sethisttrim}[2]{\expandafter\def\csname histtrim@#1\endcsname{#2}}
\newcommand{\histtrim}[1]{%
\ifcsname histtrim@#1\endcsname
  \csname histtrim@#1\endcsname
\else
  \histtrimdefault
\fi
}
\sethisttrim{guided_pibt}{1.1cm 1.3cm 0 0}
\newcommand{\splitkeyval}[1]{\splitkeyvalaux#1\relax}
\def\splitkeyvalaux#1=#2\relax{#1 & = & #2}
\newcommand{\metricstwo}[2]{%
{\renewcommand{\arraystretch}{0.9}%
\begin{tabular}{@{}l@{\,}c@{\,}l@{}}
\splitkeyval{#1}\\
\splitkeyval{#2}
\end{tabular}}%
}
\newcommand{\entrymetrics}[2]{%
\if\relax\detokenize{#1}\relax
\else
\node[anchor=north] at (2.3, 4.5) {\small \metricstwo{#1}{#2}};
\fi
}
\newcommand{\entry}[5]{
\begin{scope}[xshift=0.19*#3\linewidth, scale=0.82, transform shape]
{
\node[anchor=south west] at (0.5, -0.1)
{\includegraphics[width=0.2\linewidth]{\heatmapfile{#2}}};
\node[] at (2.3, 4.75) {\small #1};
\entrymetrics{#4}{#5}
%
% usage
\node[anchor=north west] at (0.65, -0.0)
{%
\begingroup
\edef\histtrimvals{\histtrim{#2}}%
\edef\histcmd{%
\noexpand\includegraphics[width=0.2\noexpand\linewidth,clip,trim={\histtrimvals}]{\noexpand\histogramfile{#2}}%
}%
\histcmd
\endgroup
};

\node[anchor=north east, overlay] at (0.9, -0.65){\scriptsize 0};
\node[anchor=north east, overlay] at (0.9, -0.45){\scriptsize 1};
\node[anchor=north east, overlay] at (0.9, -0.26){\scriptsize 10};
\node[anchor=north east, overlay] at (0.9, -0.07){\scriptsize 100};
\node[anchor=north, overlay] at (0.9, -0.8){\scriptsize 1};
\node[anchor=north, overlay] at (1.45, -0.8){\scriptsize 100};
\node[anchor=north, overlay] at (2.1, -0.8){\scriptsize 200};
\node[anchor=north, overlay] at (2.8, -0.8){\scriptsize 300};
\node[anchor=north, overlay] at (3.5, -0.8){\scriptsize 400};
\node[anchor=north, overlay] at (4.10, -0.8){\scriptsize 500};

\node[anchor=north east,rotate=0] at (0.9, 0.15) {\scriptsize freq.};
\node[anchor=north west] at (1.5, -1.02) {\scriptsize number of stops};
}
\end{scope}
}
\textbf{\mapname{empty-48-48}, 1,000 agents}\par\vspace{0.2em}
\begin{tikzpicture}
\entry{RHCR (failed)}{RHCR}{0}{throughput=3.07}{runtime=11.4 [s/step]}
\entry{PIBT}{pibt_tiebreaking}{1}{throughput=19.4}{runtime=0.0011 [s/step]}
\entry{LaCAM}{lifelong_lacam}{2}{throughput=21.4}{runtime=0.032 [s/step]}
\entry{Guided-PIBT}{guided_pibt}{3}{throughput=20.0}{runtime=0.013 [s/step]}
\entry{LLLG (proposed)}{lifelong_lg_lacam}{4}{throughput=\textbf{27.3}}{runtime=0.26 [s/step]}
\node[anchor=south west] at (0.95\linewidth, -0.0)
{\includegraphics[width=0.028\linewidth]{fig/raw/heatmap/colorbar_stops}};
\node[anchor=south east,rotate=-90] at (0.98\linewidth, 0.45) {\small number of stops};
\end{tikzpicture}
}
\end{minipage}
\vspace{-1.0em}
\caption{Heatmap of vertex stop counts, where congestion is implied by warm colors.
The bottom of each heatmap shows the distribution among vertices of how many times each vertex is used as a stop by any agent.
}
\label{fig:heatmap3}
\end{figure*}
}

\subsection{Qualitative Comparison}

We first visualize the effect of guidance in the lifelong setting on two dense benchmark maps: \mapname{random-64-64-10} with $1{,}000$ agents and \mapname{empty-48-48} with $1{,}000$ agents. \Cref{fig:heatmap3} compares five LMAPF solvers: RHCR, PIBT, LaCAM, Guided-PIBT, and LLLG. Their qualitative differences are reflected both in the heatmaps and in the stop-count distributions shown at the bottom of the figure.

In \mapname{random-64-64-10}, RHCR appears to achieve near-optimal throughput, but at high planning cost.
PIBT and LaCAM, lacking explicit congestion-mitigation guidance, concentrate agent interactions near the center of the map and produce pronounced local clusters with frequent stops, although LaCAM appears somewhat less congested, likely due to the hindrance tiebreaking.
Guided-PIBT generates coarse detours to mitigate congestion, which disperses stop locations across the map. However, its guidance is time-independent and tends to discard fine-grained spatiotemporal trajectory information, making it overly conservative and leaving avoidable waiting in local bottlenecks.
In contrast, LLLG produces informative spatiotemporal local guidance precisely where agents become locally dense, smoothing traffic in bottleneck regions and reducing repeated stopping.
Notably, it achieves slightly higher throughput than RHCR while planning about ten times faster.

In \mapname{empty-48-48}, RHCR fails frequently in this dense setting.
A key reason is that its PBS-based planning starts from an infeasible state, and a feasible solution may not be found within the allotted time limit.
In such cases, RHCR falls back to LRA$^*$~\cite{silver2005cooperative}, which often results in excessive waiting and a marked drop in throughput.
As more agents remain stationary, congestion further worsens, and once the number of non-moving agents exceeds half of the team, the episode is declared a failure and terminated.
For the other methods, the qualitative trends become even clearer: PIBT and LaCAM show concentration and stopping, Guided-PIBT disperses traffic more broadly yet still leaves residual waiting, and LLLG most effectively smooths traffic and suppresses repeated stops.

\subsection{Quantitative Comparison}
\Cref{fig:highlight} presents comparisons of lifelong methods across four representative scenarios. Among them, PIBT-based solvers (PIBT, LaCAM, WPPL, and Guided-PIBT) compute feasible actions instantly.
RHCR, in contrast, is built on PBS and is designed to produce promising suboptimal plans in relatively sparse environments, but it incurs substantially higher planning time. 
Moreover, in \mapname{ht_chantry}, the solution quality can lag behind that of PIBT-based solvers. 
LLLG inherits the computational efficiency of PIBT-based planning while producing consistently better solutions than RHCR, demonstrating that local guidance can bridge the gap between speed and quality.

\Cref{fig:result-main} provides a comprehensive comparison across maps and agent scales. 
In sparse scenarios (e.g., $400$ agents in \mapname{random-64-64-20}), LLLG achieves solution quality comparable to RHCR while requiring much less runtime. In dense scenarios (e.g., $1{,}000$ agents in \mapname{random-64-64-20}), LLLG substantially outperforms prior methods in throughput, indicating that local guidance is particularly effective when congestion dominates system performance.

An exception is the large \mapname{warehouse-20-40-10-2-1}, where Guided-PIBT is particularly effective.
We conjecture that this is because the guidance window cannot be chosen to cover the full length of long corridors.
As a result, local finite-window guidance can struggle on maps containing one-cell-wide corridors, where agents must account in advance for oncoming traffic that may still lie beyond the current window.
This observation is aligned with local guidance for one-shot MAPF~\cite{arita2025local}.

{
\setlength{\tabcolsep}{1pt}
\newcommand{\entryheadheight}{4.8em}
\newcommand{\mapplotsep}{0.0em}
\newcommand{\xlabellegendsep}{0.2em}
\newcommand{\maplabel}[1]{%
  \ifthenelse{\equal{#1}{warehouse-10-20-10-2-2}}{%
    \mapname{warehouse}\\[-0.45em]
    \mapname{10-20-10-2-2}%
  }{%
    \ifthenelse{\equal{#1}{warehouse-20-40-10-2-1}}{%
      \mapname{warehouse}\\[-0.45em]
      \mapname{20-40-10-2-1}%
    }{%
      \ifthenelse{\equal{#1}{random-32-32-10}}{%
        \mapname{random}\\[-0.45em]
        \mapname{32-32-10}%
      }{%
        \ifthenelse{\equal{#1}{random-64-64-20}}{%
          \mapname{random}\\[-0.45em]
          \mapname{64-64-20}%
        }{%
          \mapname{#1}%
        }%
      }%
    }%
  }%
}
\newcommand{\mapentry}[3]{%
  \begin{minipage}[t]{0.12\linewidth}
    \centering
    \setlength{\baselineskip}{0.75\baselineskip}
    \scriptsize
    \parbox[t][\entryheadheight][c]{\linewidth}{%
      \begin{minipage}[c]{0.28\linewidth}
        \centering
        \includegraphics[width=\linewidth]{fig/raw/maps/#1}
      \end{minipage}\hspace{-0.02\linewidth}%
      \begin{minipage}[c]{0.7\linewidth}
        \centering
        {\tiny\maplabel{#1}}\\[-0.2em]
        {\tiny\makebox[\linewidth][c]{(#2)}}%
      \end{minipage}%
    }%
  \end{minipage}%
}
\newcommand{\plotentry}[2]{%
  \begin{minipage}[t]{0.12\linewidth}
    \centering
    \includegraphics[width=\linewidth]{fig/raw/main_results/#1_#2}
  \end{minipage}%
}
\newcommand{\ylabeltp}{\rotatebox{90}{\small \hspace{0.3em}{ throughput $\rightarrow$}}}
\newcommand{\ylabelrt}{\rotatebox{90}{\small \hspace{0.3em}{  $\leftarrow$ runtime [\SI{}{\second}]}}}
\newcommand{\xlabel}{\small agents}
\begin{figure*}[t!]
  \centering
  \begin{tabular}{@{}c@{}cccccccc@{}}
    &
    \mapentry{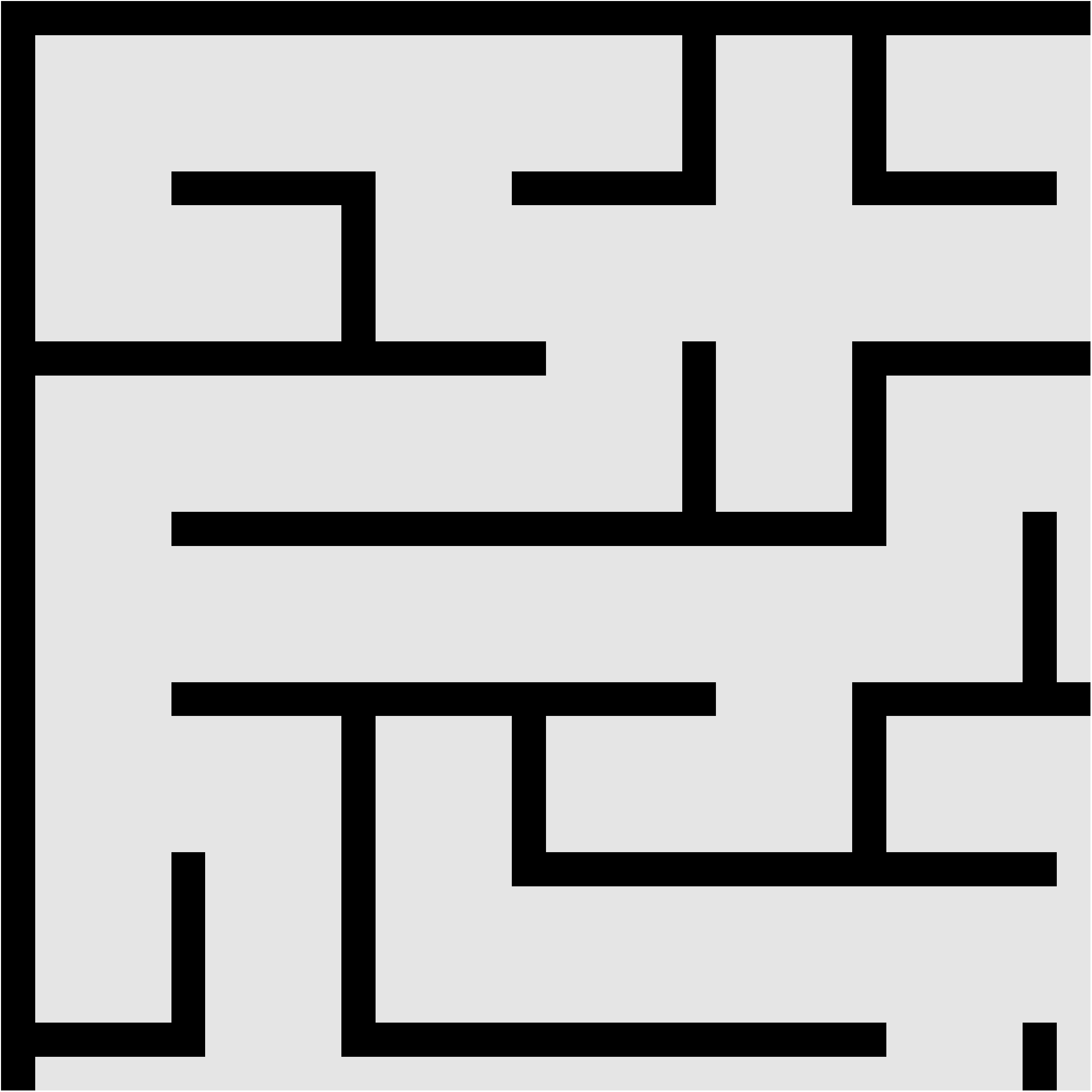}{32x32}{790} &
    \mapentry{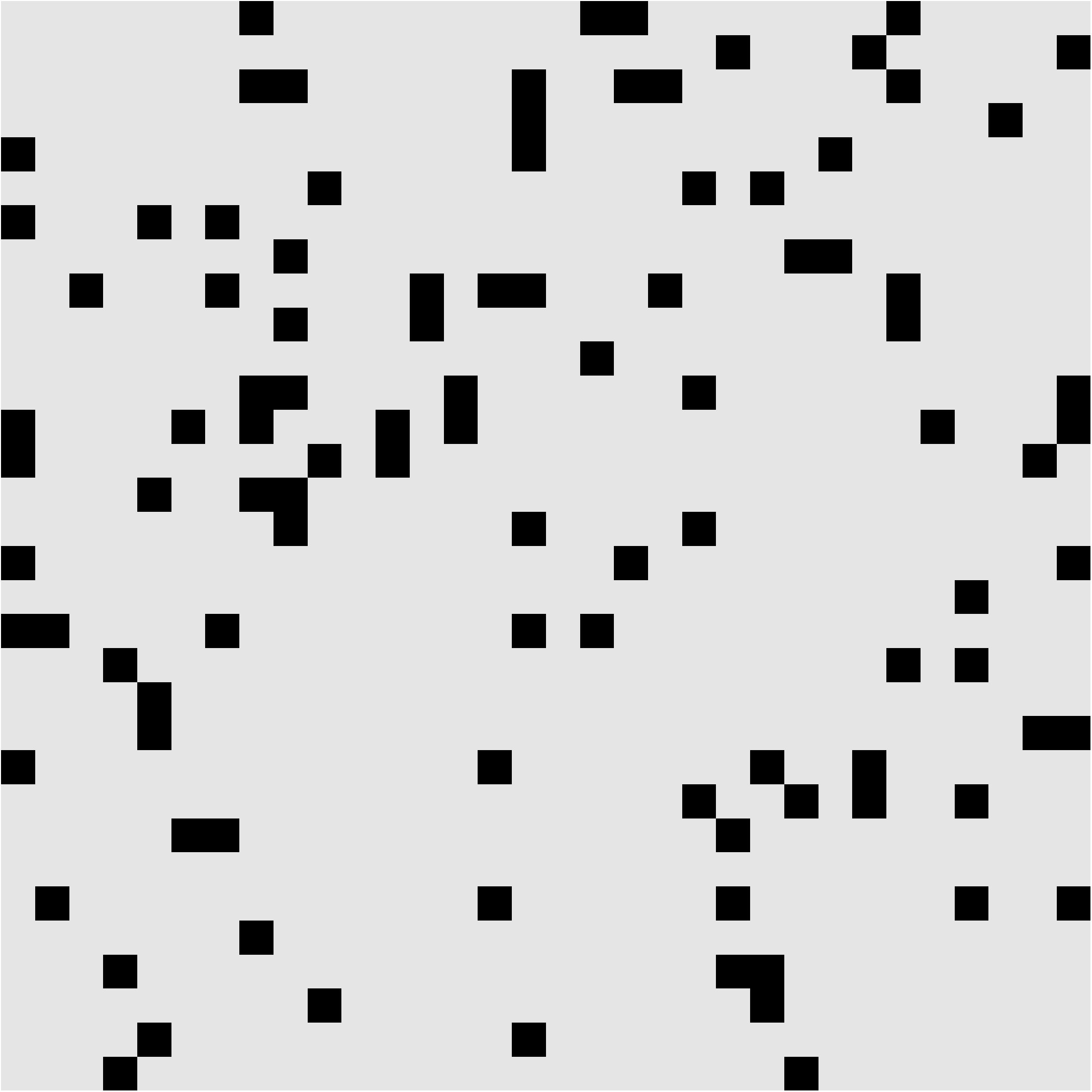}{32x32}{922} &
    \mapentry{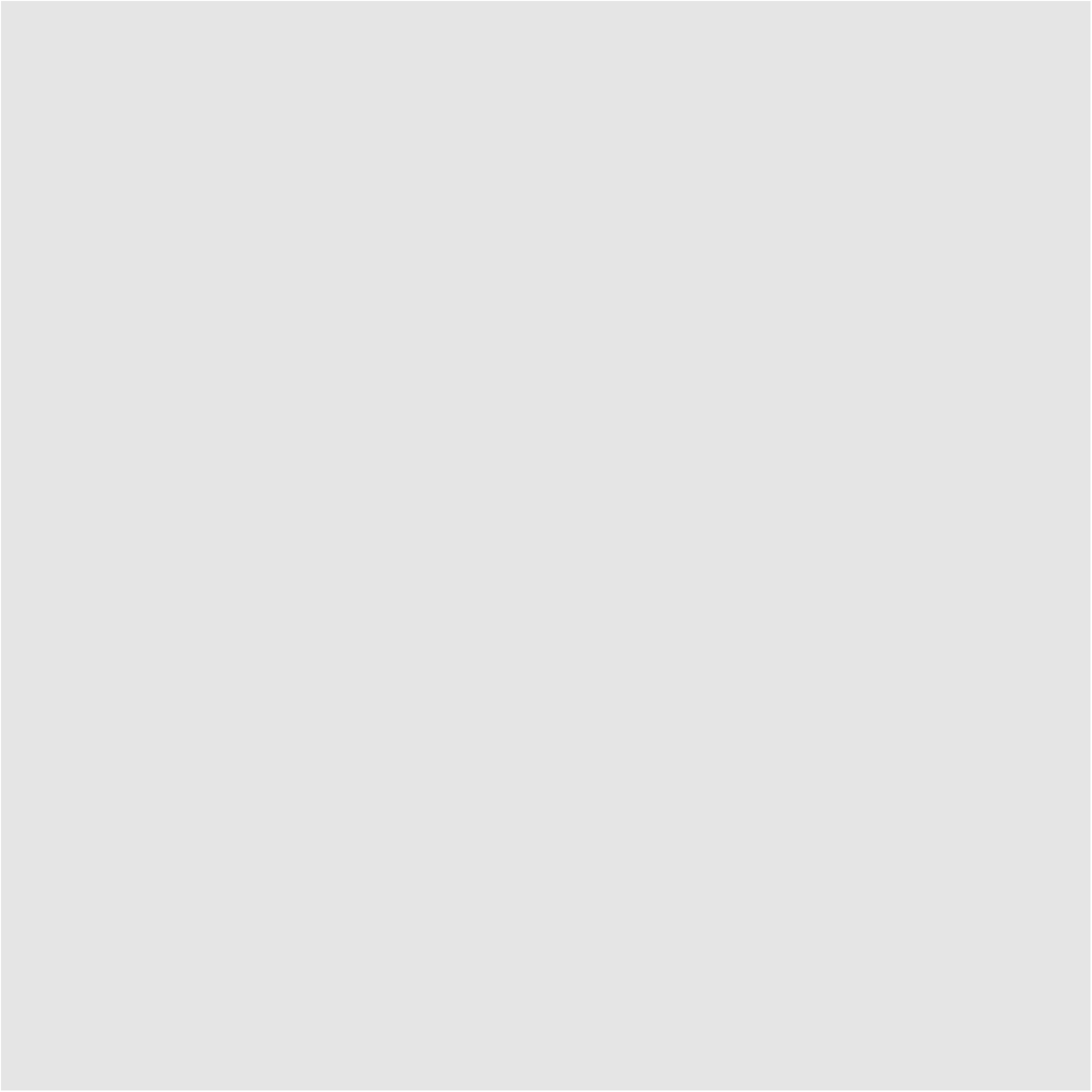}{48x48}{2,304} &
    \mapentry{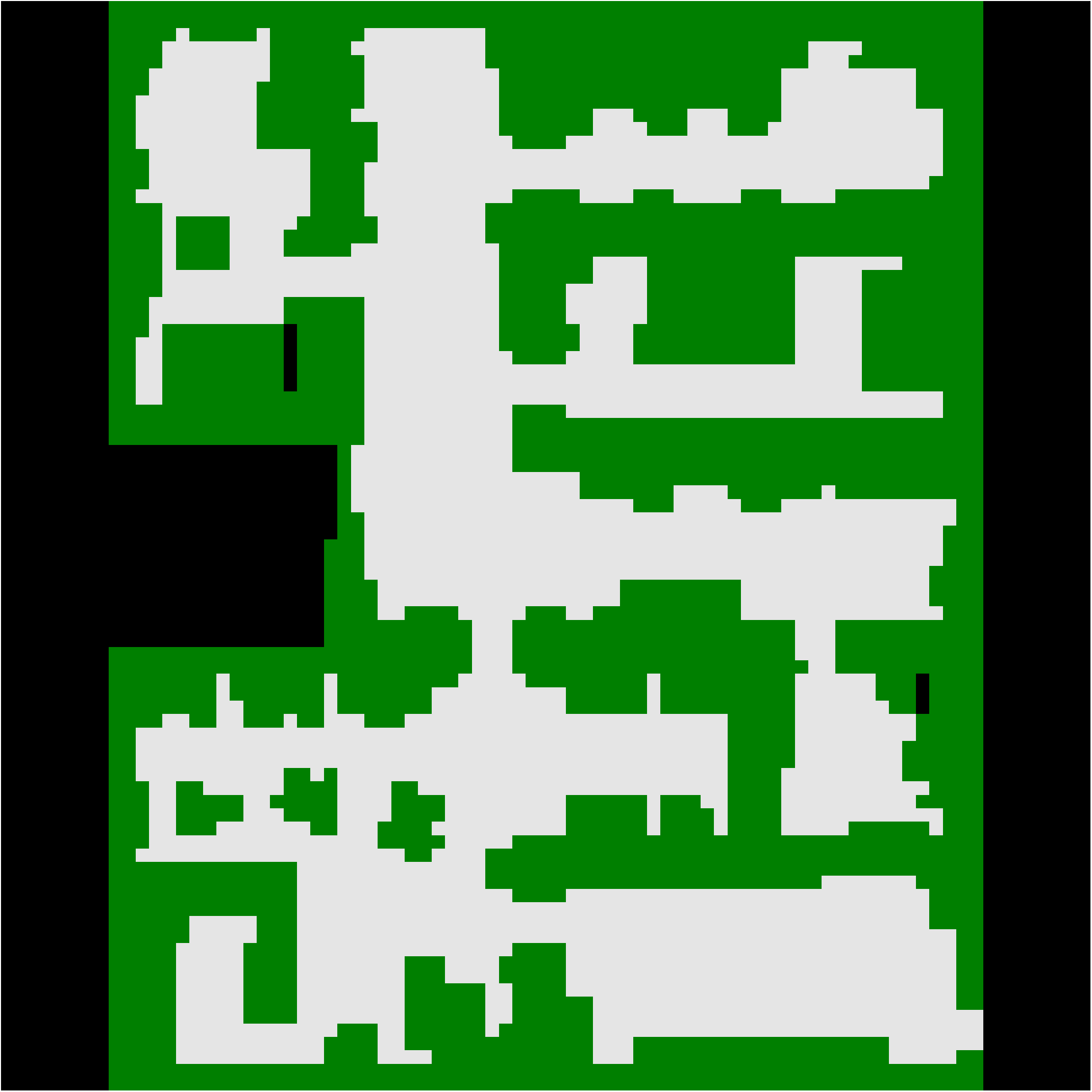}{65x81}{2,445} &
    \mapentry{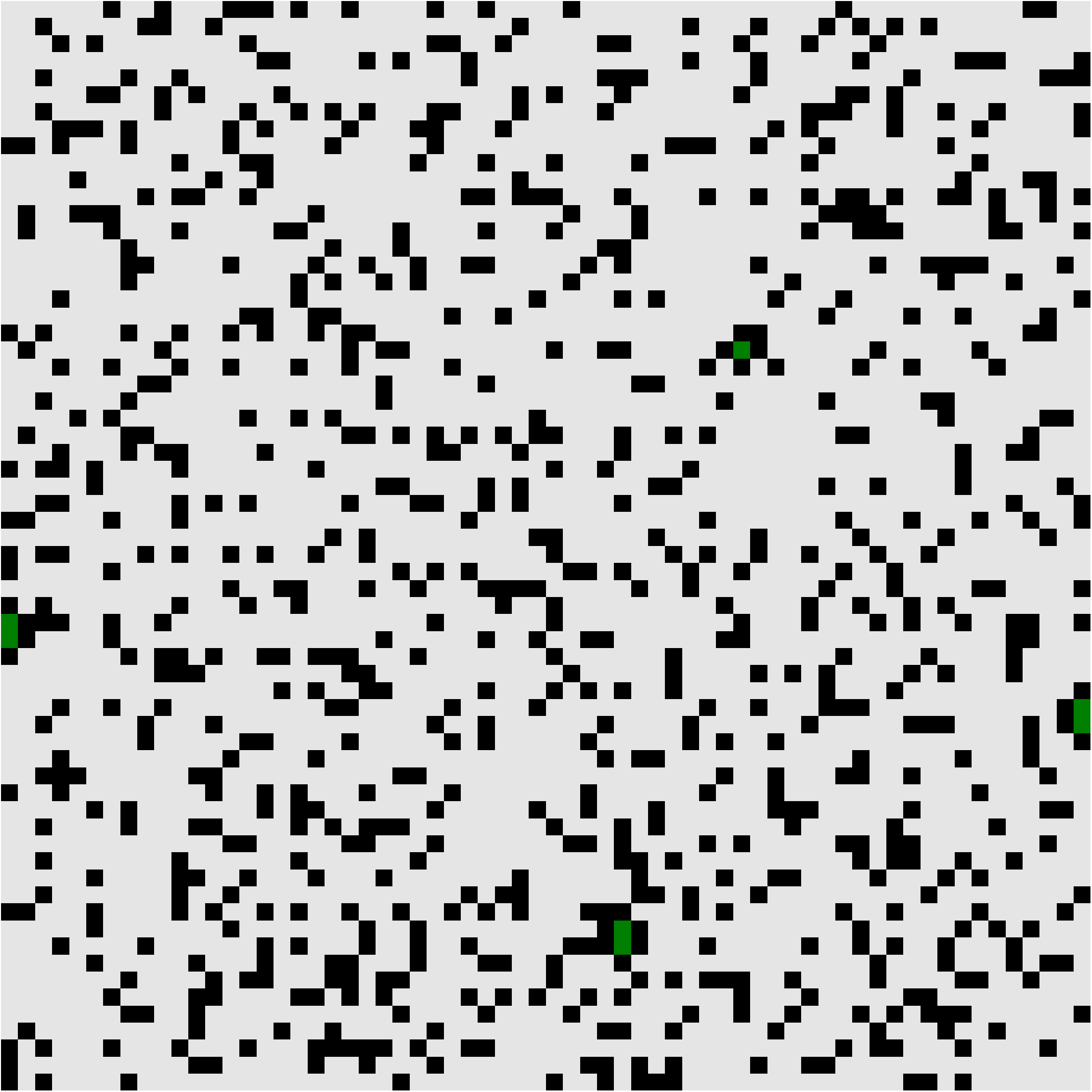}{64x64}{3,687} &
    % \mapentry{maze-128-128-10}{128x128}{14,818} &
    \mapentry{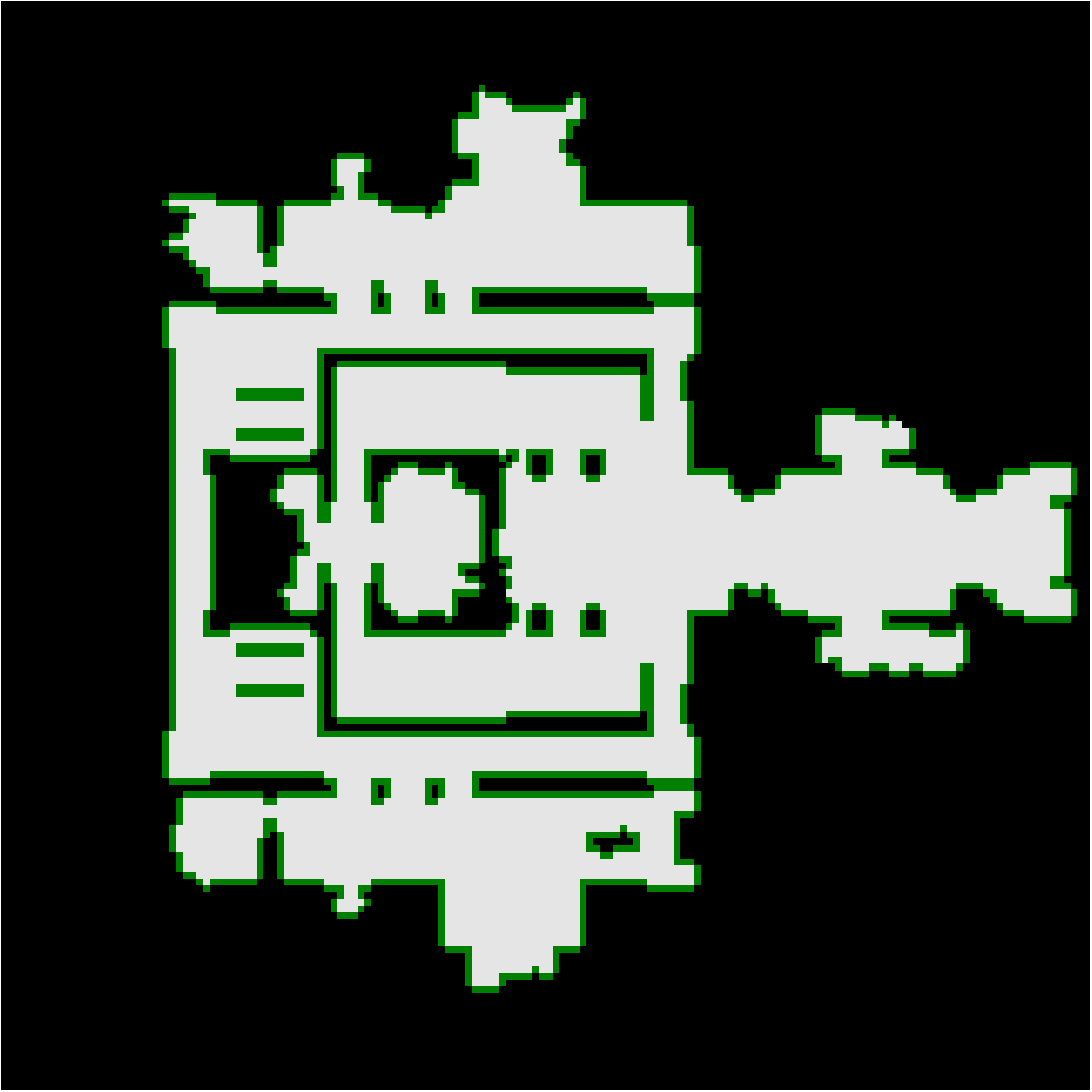}{162x141}{7,461} &
    % \mapentry{ost003d}{194x194}{13,214} &
    % \mapentry{lak303d}{194x194}{14,784} &
    % \mapentry{Boston_0_256}{256x256}{47,768} &
    % \mapentry{Paris_1_256}{256x256}{47,768} &
    \mapentry{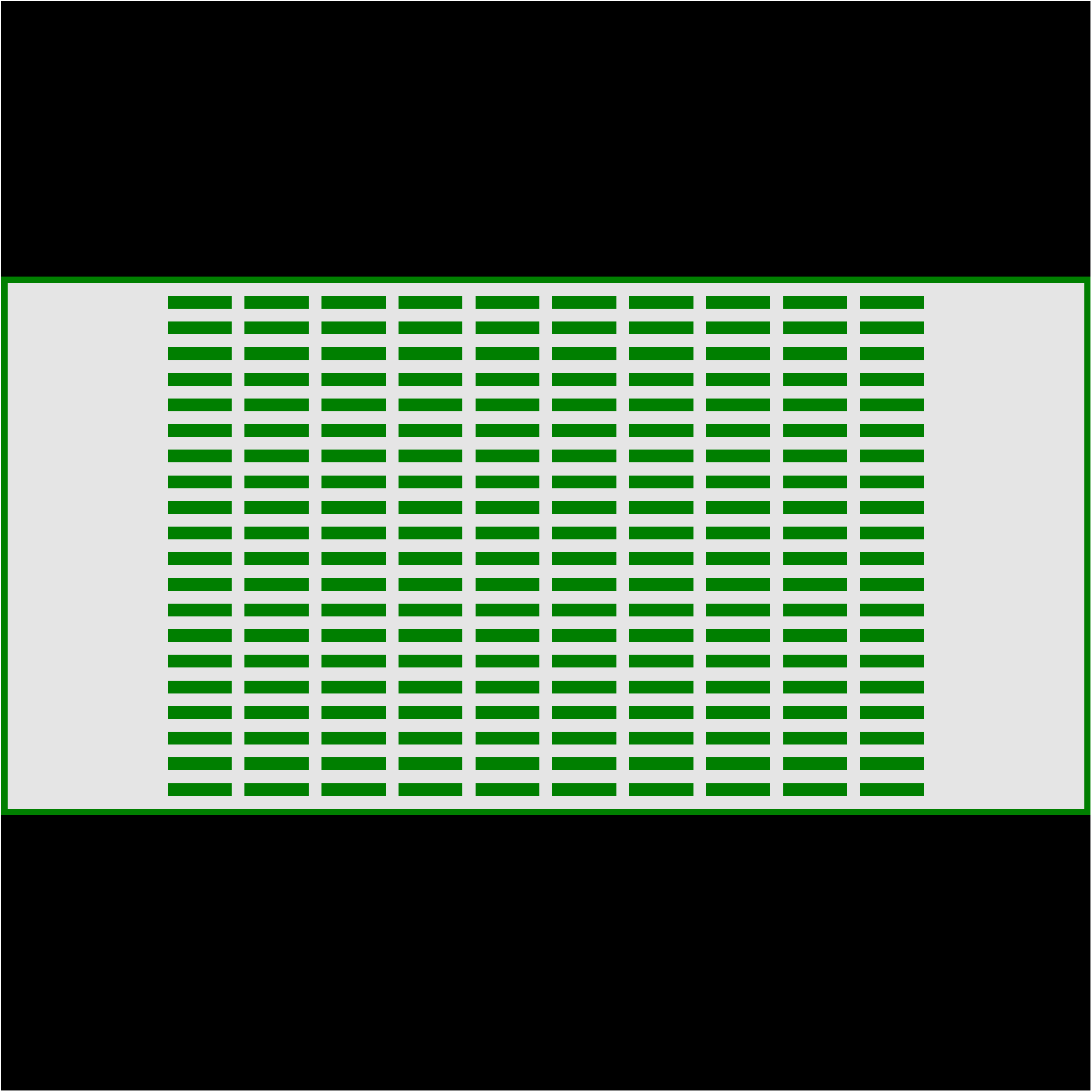}{170x84}{9,776} &
    \mapentry{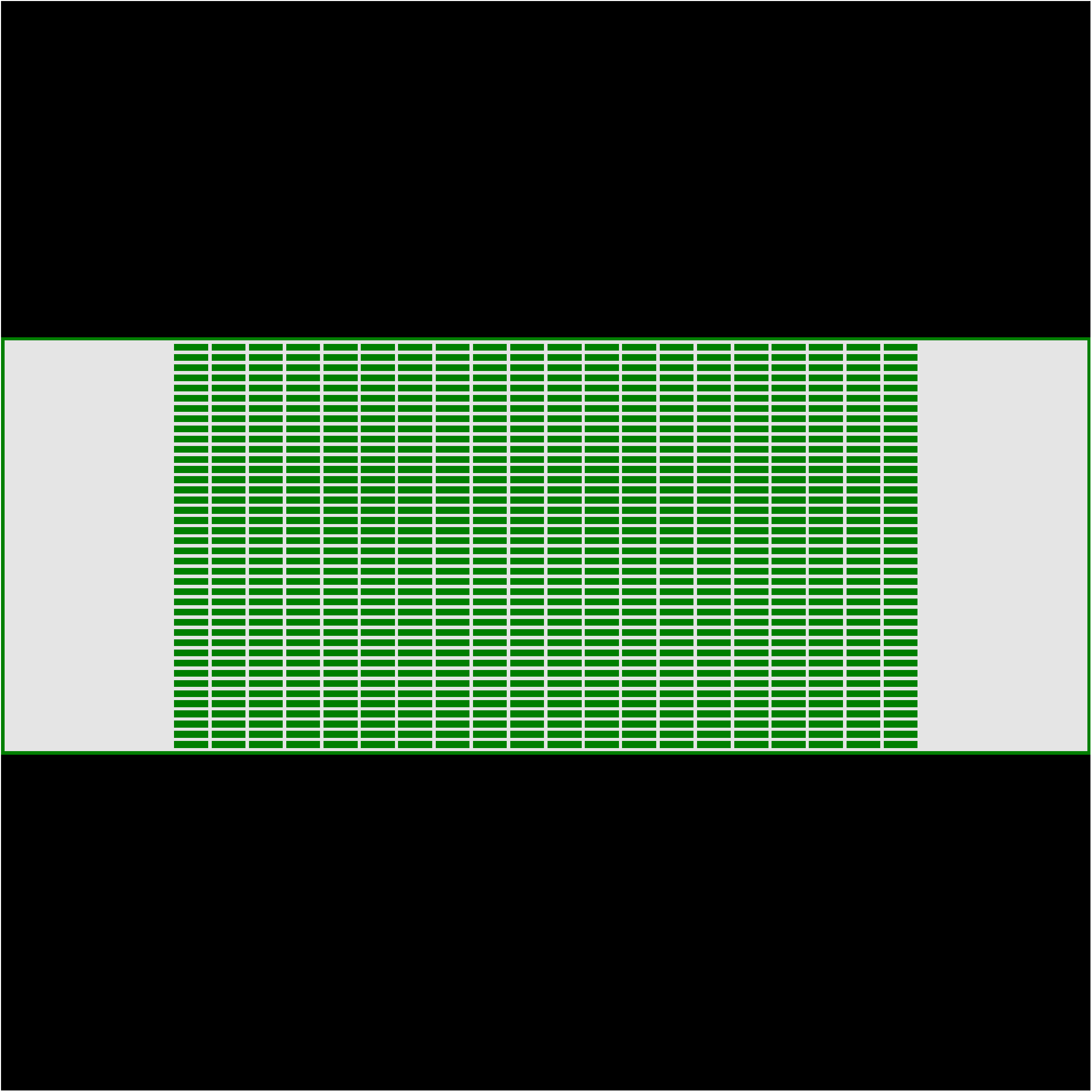}{321x123}{22,599}
    \\[0.8em]
    \noalign{\vskip\mapplotsep}
    \ylabeltp &
    \plotentry{agents_throughput}{maze-32-32-4} &
    \plotentry{agents_throughput}{random-32-32-10} &
    \plotentry{agents_throughput}{empty-48-48} &
    \plotentry{agents_throughput}{den312d} &
    \plotentry{agents_throughput}{random-64-64-20} &
    % \plotentry{agents_throughput}{maze-128-128-10} &
    \plotentry{agents_throughput}{ht_chantry} &
    % \plotentry{agents_throughput}{ost003d} &
    % \plotentry{agents_throughput}{lak303d} &
    % \plotentry{agents_throughput}{Boston_0_256} &
    % \plotentry{agents_throughput}{Paris_1_256} &
    \plotentry{agents_throughput}{warehouse-10-20-10-2-2} &
    \plotentry{agents_throughput}{warehouse-20-40-10-2-1}
    \\[0.3em]
    \ylabelrt &
    \plotentry{agents_runtime}{maze-32-32-4} &
    \plotentry{agents_runtime}{random-32-32-10} &
    \plotentry{agents_runtime}{empty-48-48} &
    \plotentry{agents_runtime}{den312d} &
    \plotentry{agents_runtime}{random-64-64-20} &
    % \plotentry{agents_runtime}{maze-128-128-10} &
    \plotentry{agents_runtime}{ht_chantry} &
    % \plotentry{agents_runtime}{ost003d} &
    % \plotentry{agents_runtime}{lak303d} &
    % \plotentry{agents_runtime}{Boston_0_256} &
    % \plotentry{agents_runtime}{Paris_1_256} &
    \plotentry{agents_runtime}{warehouse-10-20-10-2-2} &
    \plotentry{agents_runtime}{warehouse-20-40-10-2-1}
    \\
    & \multicolumn{8}{c}{\xlabel}
    \\
    \noalign{\vskip\xlabellegendsep}
    & \multicolumn{8}{c}{\includegraphics[width=0.95\linewidth,clip,trim={0 1.6cm 0 1.3cm}]{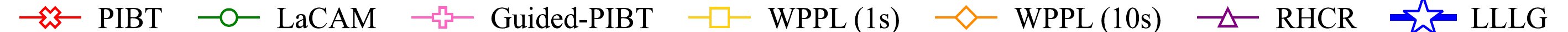}}
    \\
  \end{tabular}
  \caption{
  Performance evaluation of lifelong MAPF methods, assessed by throughput and runtime.
  For each map, the two subfigures report throughput and runtime over 500 steps for each agent scale.
  Each subfigure reports the average over the solved instances among 10 cases per setting, and the shaded regions indicate the min/max values.
  Results for failed cases (RHCR) are plotted with dashed lines, where each value is computed from the period before failure.
  % The size for each map is shown in parentheses.
  }
  \label{fig:result-main}
\end{figure*}
}

{
\begin{figure}[t!]
\centering
\begin{tikzpicture}
\node[anchor=south] at (0, 0)
{\includegraphics[width=0.45\linewidth]{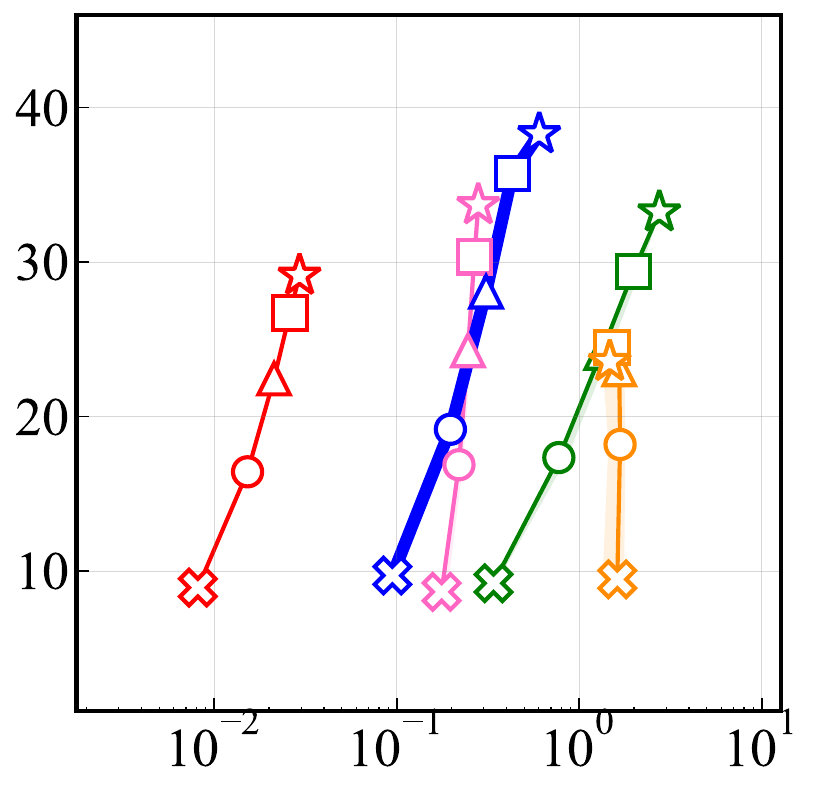}};
\node[] at (0, 0) {\small $\leftarrow$ runtime [\SI{}{\second}]};
\node[rotate=90] at (-0.27\linewidth, 0.25\linewidth) {\small throughput $\rightarrow$};
\node[anchor=west] at (0.55, 3.05) {\small \textcolor[HTML]{008000}{LaCAM}};
\node[anchor=west] at (0.49, 0.8)  {\small \textcolor[HTML]{FF8C00}{WPPL}};
\node[anchor=west] at (0.10, 3.5)  {\small \textcolor[HTML]{0000FF}{LLLG}};
\node[anchor=west] at (-1.55, 2.2)  {\small \textcolor[HTML]{FF0000}{PIBT}};
\node[anchor=west] at (-0.9, 3.1)  {\small \textcolor[HTML]{ff66c4}{\shortstack[c]{Guided-\\PIBT}}};

\begin{scope}[xshift=0.27\linewidth,yshift=0.15\linewidth]
  \draw[] (0, 0) --
  node[pos=0]{\small $\times$}
  node[pos=0.25]{$\bullet$}
  node[pos=0.5]{$\blacktriangle$}
  node[pos=0.75]{\scriptsize $\blacksquare$}
  node[pos=1.0]{\scriptsize $\bigstar$}
  node[right,pos=0]{\footnotesize 2,000}
  node[right,pos=0.25]{\footnotesize 4,000}
  node[right,pos=0.5]{\footnotesize 6,000}
  node[right,pos=0.75]{\footnotesize 8,000}
  node[right,pos=1]{\footnotesize 10,000}
  node[above right=0.5,pos=1.05]{\footnotesize agents}
  (0.2, 1.8);
\end{scope}
\end{tikzpicture}
\caption{
  Scalability test on \mapname{warehouse-20-40-10-2-2}. LLLG uses a lightweight setting ($w_{\Phi}{=}8$, $w_{\Pi}{=}5$).
}
\label{fig:scalability}
\end{figure}
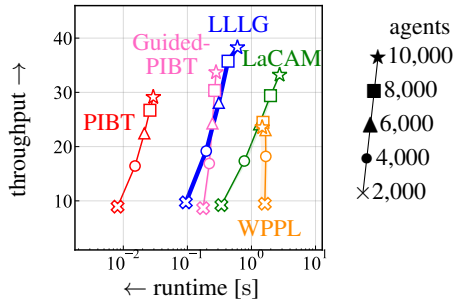
}

\paragraph{Scalability.}
\Cref{fig:scalability} evaluates scalability in LMAPF on the warehouse map with up to $10{,}000$ agents.
Given its problem scale, LLLG uses lightweight parameters: $w_{\Phi}{=}8$ and $w_{\Pi}{=}5$.
In this setting, RHCR fails to keep up with $2 {,}000$ agents, as repeated branch-and-replan becomes a computational bottleneck.
In contrast, LLLG remains practical at an ultra-large scale.
Across the tested loads, LLLG consistently produces actions within \SI{1}{\second} per step even for $10{,}000$ agents, and achieves approximately a 30\% throughput gain over PIBT.
These results indicate that LLLG scales effectively to dense lifelong operation.

\subsection{Design Justification}
To justify the design of our LLLG framework and better understand its empirical advantage, we isolate its main components and examine their contributions to performance.
This section uses the shorter LMAPF simulation steps (100).

\paragraph{Warm Start with Previous-step Solution.}
\Cref{fig:realtime} evaluates the effect of warm-starting in LLLG.
As a baseline, we include LLLG$_{\boldsymbol{\emptyset}}$, which reconstructs guidance from scratch at every timestep.
LLLG$_{\boldsymbol{\Phi}}$ inherits the guidance paths from the previous guidance, whereas LLLG$_{\boldsymbol{\Pi}}$ initializes guidance from the remaining suffix of the previous finite-horizon solution (i.e., equivalent to LLLG previously evaluated).
As shown in \cref{fig:realtime}, LLLG$_{\boldsymbol{\Phi}}$ clearly improves throughput over LLLG$_{\boldsymbol{\emptyset}}$ at essentially the same runtime, and LLLG$_{\boldsymbol{\Pi}}$ yields a further improvement without additional computation cost.
This suggests that warm-starting local guidance is beneficial in LMAPF and that reusing the previous-step solution is more effective than inheriting guidance alone.
We speculate that guidance alone provides only a soft bias with respect to collisions, whereas the previous-step solution gives an explicit collision-free forecast over the near future and therefore a stronger initialization for subsequent search.

\paragraph{Horizon Length and Guidance Window.}
\Cref{fig:window} illustrates the interaction between the planning horizon length $w_{\Pi}$ and the guidance window size $w_{\Phi}$ used by LLLG across multiple scenarios.
For most agent configurations, increasing the planning horizon improves throughput up to a broad plateau, indicating that horizon planning is more effective when it is long enough to resolve near-term interactions and provide informative guidance.
This appears consistent with the LLLG$_{\boldsymbol{\Phi}}$ versus LLLG$_{\boldsymbol{\Pi}}$ comparison in \cref{fig:realtime}, where guidance constructed from a collision-resolving horizon plan tends to coordinate agents more effectively than guidance alone.
At the same time, overly long horizons (e.g., $w_{\Pi}{=}20$) become less useful because future goal assignments are unknown in LMAPF;
additional planning effort is spent on predictions that will soon be invalidated, making the added computation cost less cost-effective.

{
\setlength{\tabcolsep}{2pt}
\newcommand{\realtimefig}[1]{fig/raw/realtime/runtime_throughput_#1}

% method label positions (x,y) in normalized image coordinates
\newcommand{\methodposSmallLaCAM}{{0.15,0.12}}
\newcommand{\methodposSmallLG}{{0.47,0.92}}
\newcommand{\methodposSmallWPPL}{{0.47,0.84}}
\newcommand{\methodposSmallRHCR}{{0.42,0.45}}
\newcommand{\methodposSmallGuidedPIBT}{{0.27,0.29}}

\newcommand{\methodposLargeLaCAM}{{0.15,0.12}}
\newcommand{\methodposLargeLG}{{0.6,0.92}}
\newcommand{\methodposLargeWPPL}{{0.6,0.76}}
\newcommand{\methodposLargeRHCR}{{0.42,0.26}}
\newcommand{\methodposLargeGuidedPIBT}{{0.42,0.51}}

\newcommand{\WPPLLabelSmall}{LLLG$_{\boldsymbol{\Phi}}$}
\newcommand{\WPPLLabelLarge}{LLLG$_{\boldsymbol{\Phi}}$}

\newcommand{\methodlabel}[5]{% #1: panel key, #2: method key, #3: color, #4: label, #5: anchor
  \node[anchor=#5] at (\csname methodpos#1#2\endcsname) {\scriptsize \textcolor[HTML]{#3}{\shortstack[c]{#4}}};
}
\newcommand{\methodlabels}[1]{% #1: panel key
  \methodlabel{#1}{LaCAM}{2ca02c}{LaCAM}{west}
  \methodlabel{#1}{LG}{1f77b4}{LG}{west}
  \methodlabel{#1}{WPPL}{e377c2}{WPPL}{west}
  \methodlabel{#1}{RHCR}{9467bd}{RHCR}{west}
  \methodlabel{#1}{GuidedPIBT}{ff7f0e}{Guided-PIBT}{west}
}

\newcommand{\ylabel}{\rotatebox{90}{\small \hspace{3.0em}{ throughput $\rightarrow$}}}
\newcommand{\xlabel}{\small  $\leftarrow$ runtime [\SI{}{\second}]}

\newcommand{\agentlegendSmall}{% [100, 200, 300, 400]
  \begin{scope}[shift={(0.7,0.15)}]
    \draw[] (0, 0) --
    node[pos=0]{\tiny $\times$}
    node[pos=0.333]{\tiny $\bullet$}
    node[pos=0.666]{\tiny $\blacktriangle$}
    node[pos=1.0]{\tiny $\blacksquare$}
    node[right,pos=0]{\tiny 100}
    node[right,pos=0.333]{\tiny 200}
    node[right,pos=0.666]{\tiny 300}
    node[right,pos=1]{\tiny 400}
    node[above right=0.2,pos=1]{\tiny agents}
    (0.06, 0.24);
  \end{scope}%
}
\newcommand{\agentlegendLarge}{% [200, 400, 600, 800, 1000]
  \begin{scope}[shift={(0.7,0.15)}]
    \draw[] (0, 0) --
    node[pos=0]{\tiny $\times$}
    node[pos=0.25]{\tiny $\bullet$}
    node[pos=0.5]{\tiny $\blacktriangle$}
    node[pos=0.75]{\tiny $\blacksquare$}
    node[pos=1.0]{\tiny $\bigstar$}
    node[right,pos=0]{\tiny 200}
    node[right,pos=0.25]{\tiny 400}
    node[right,pos=0.5]{\tiny 600}
    node[right,pos=0.75]{\tiny 800}
    node[right,pos=1]{\tiny 1000}
    node[above right=0.2,pos=1]{\tiny agents}
    (0.06, 0.24);
  \end{scope}%
}

\newcommand{\entry}[2]{% #1: instance key, #2: panel key
  \begin{minipage}[t]{0.38\linewidth}
    \centering
    \begin{tikzpicture}
      \node[anchor=south west,inner sep=0] (img) at (0, 0) {%
        \includegraphics[width=\linewidth]{\realtimefig{#1}}%
      };
      \begin{scope}[x={(img.south east)},y={(img.north west)}]
        % draw labels per-figure
        % \methodlabel{#2}{LaCAM}{2ca02c}{LaCAM}{west}
        \methodlabel{#2}{LG}{0000FF}{LLLG$_{\boldsymbol{\Pi}}$}{west}
        \methodlabel{#2}{WPPL}{008000}{\csname WPPLLabel#2\endcsname}{west}
        \methodlabel{#2}{RHCR}{800080}{RHCR}{west}
        \methodlabel{#2}{GuidedPIBT}{FF8C00}{LLLG$_{\boldsymbol{\emptyset}}$}{west}
        \csname agentlegend#2\endcsname
      \end{scope}
    \end{tikzpicture}
  \end{minipage}%
}
\begin{figure}[t!]
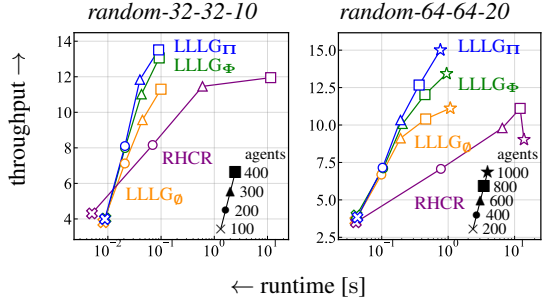

  \centering
  \begin{tabular}{@{}c@{\hspace{0.8em}}cc@{}}
    & \small \mapname{random-32-32-10} & \small \mapname{random-64-64-20} \\
    \ylabel &
    \entry{random-32-32-10}{Small} &
    \entry{random-64-64-20}{Large} \\
    & \multicolumn{2}{c}{\xlabel} \\
  \end{tabular}
  \caption{
    % Windowed Planning, Incremental Search, Receding Horizon
    Comparison of warm-start schemes for LLLG.
  }
  \label{fig:realtime}
\end{figure}
}

{
\setlength{\tabcolsep}{2pt}
\newcommand{\windowfig}[2]{fig/raw/window2/plot_lg_window_lacam_horizon_collab_agents#2_#1}

% window label positions (x,y) in normalized image coordinates (per panel)
\newcommand{\labelposWindowSmallA}{{0.16,0.23}}
\newcommand{\labelposWindowSmallB}{{0.3,0.6}}
\newcommand{\labelposWindowSmallC}{{0.43,0.73}}
\newcommand{\labelposWindowSmallD}{{0.69,0.73}}

\newcommand{\labelposWindowLargeA}{{0.25,0.23}}
\newcommand{\labelposWindowLargeB}{{0.21,0.48}}
\newcommand{\labelposWindowLargeC}{{0.4,0.78}}
\newcommand{\labelposWindowLargeD}{{0.68,0.78}}

\newcommand{\labelposWindowMediumA}{{0.2,0.23}}
\newcommand{\labelposWindowMediumB}{{0.3,0.48}}
\newcommand{\labelposWindowMediumC}{{0.47,0.6}}
\newcommand{\labelposWindowMediumD}{{0.65,0.73}}

\newcommand{\windowlabel}[5]{% #1: panel key, #2: label key, #3: color, #4: label, #5: anchor
  \node[anchor=#5] at (\csname labelposWindow#1#2\endcsname) {\scriptsize \textcolor[HTML]{#3}{\textbf{#4}}};
}
\newcommand{\windowlabels}[1]{% #1: panel key
  \windowlabel{#1}{A}{ff7f0e}{$w_{\Phi}$=3}{west}
  \windowlabel{#1}{B}{2ca02c}{$w_{\Phi}$=8}{west}
  \windowlabel{#1}{C}{d62728}{$w_{\Phi}$=15}{west}
  \windowlabel{#1}{D}{1f77b4}{$w_{\Phi}$=20}{west}
}
\newcommand{\agentlegendLarge}{% [200, 400, 600, 800, 1000]
  \begin{scope}[shift={(0.6,0.2)}]
    \draw[] (0, 0) --
    node[pos=0]{\tiny $\times$}
    node[pos=0.25]{\tiny $\bullet$}
    node[pos=0.5]{\tiny $\blacktriangle$}
    node[pos=0.75]{\tiny $\blacksquare$}
    node[pos=1.0]{\tiny $\bigstar$}
    node[right,pos=0]{\tiny 1}
    node[right,pos=0.25]{\tiny 3}
    node[right,pos=0.5]{\tiny 10}
    node[right,pos=0.75]{\tiny 20}
    node[right,pos=1]{\tiny $w_{\Pi}$=30}
    (0.04, 0.35);
  \end{scope}%
}
\newcommand{\agentlegendSmall}{%
  % \agentlegendLarge%
}
\newcommand{\agentlegendMedium}{%
  % \agentlegendLarge%
}
\newcommand{\entry}[4][0pt]{% #1: extra top padding before image, #2: map, #3: agents, #4: panel key
  \begin{minipage}[t]{0.38\linewidth}
    \centering
    {\scriptsize \mapname{#2}, #3 agents}\par
    % {\scriptsize }\par
    \vspace{0.2em}
    \vspace*{#1}
    \begin{tikzpicture}
      \node[anchor=south west,inner sep=0] (img) at (0, 0) {%
        \includegraphics[width=\linewidth]{\windowfig{#2}{#3}}%
      };
      \begin{scope}[x={(img.south east)},y={(img.north west)}]
        \windowlabels{#4}
        \csname agentlegend#4\endcsname
      \end{scope}
    \end{tikzpicture}
  \end{minipage}%
}
\newcommand{\ylabel}{\rotatebox{90}{\small \hspace{-10.5em}{ throughput $\rightarrow$}}}
\newcommand{\xlabel}{\small  $\leftarrow$ runtime [\SI{}{\second}]}
\begin{figure}[t!]
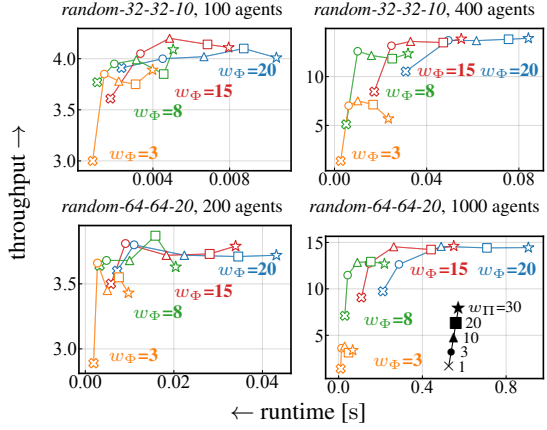

  \centering
  \begin{tabular}{@{}c@{\hspace{0.8em}}cc@{}}
    \ylabel &
    \entry{random-32-32-10}{100}{Medium} &
    \entry{random-32-32-10}{400}{Small} \\
    &
    \entry{random-64-64-20}{200}{Medium} &
    \entry[4pt]{random-64-64-20}{1000}{Large} \\
    & \multicolumn{2}{c}{\xlabel} \\
  \end{tabular}
\caption{
Sensitivity of throughput and runtime to the guidance window size $w_{\Phi}$ and the planning horizon $w_{\Pi}$.
}
\label{fig:window}
\end{figure}
}

\paragraph{Guidance Refinement.}
We next investigate guidance refinement during the construction of the predicted horizon plan at each step. In LMAPF, each agent's guidance depends on the others, and traffic pattern changes as tasks are updated online. 
To cope with this, \cref{alg:lifelong-lacam} refines guidance by repeating the guidance construction for $m$ rounds within each timestep, reusing the previous step's guidance as a cache.
\Cref{fig:refinement} shows that less frequent guidance updates are ineffective. 
In particular, $m{=}0$, which does not refine the inherited guidance and instead regenerates guidance only every window step, substantially degrades throughput. 
Updating and refining the guidance online at each step yields consistently higher throughput, indicating that ``live'' guidance based on up-to-date observations is essential for sustaining performance over long horizons.
This aligns with observations in the one-shot MAPF case~\cite{arita2025local}.

{
\setlength{\tabcolsep}{6pt}
\newcommand{\entry}[2]{%
  \begin{minipage}[t]{0.4\linewidth}
    \centering
    \includegraphics[width=#2\linewidth]{fig/raw/refinement/plot_lg_window_lg_num_refine_#1}
  \end{minipage}%
}
\newcommand{\ylabel}{\rotatebox{90}{\small \hspace{2.0em}{ throughput $\rightarrow$}}}
\newcommand{\xlabel}{\small window size}
% \begin{figure}[t!]
\begin{figure}[h]
  \centering
  \begin{tabular}{@{}c@{\hspace{0.8em}}cc@{}}
    & {\small \mapname{empty-48-48}} & {\small \mapname{random-32-32-10}} \\
    \ylabel & \entry{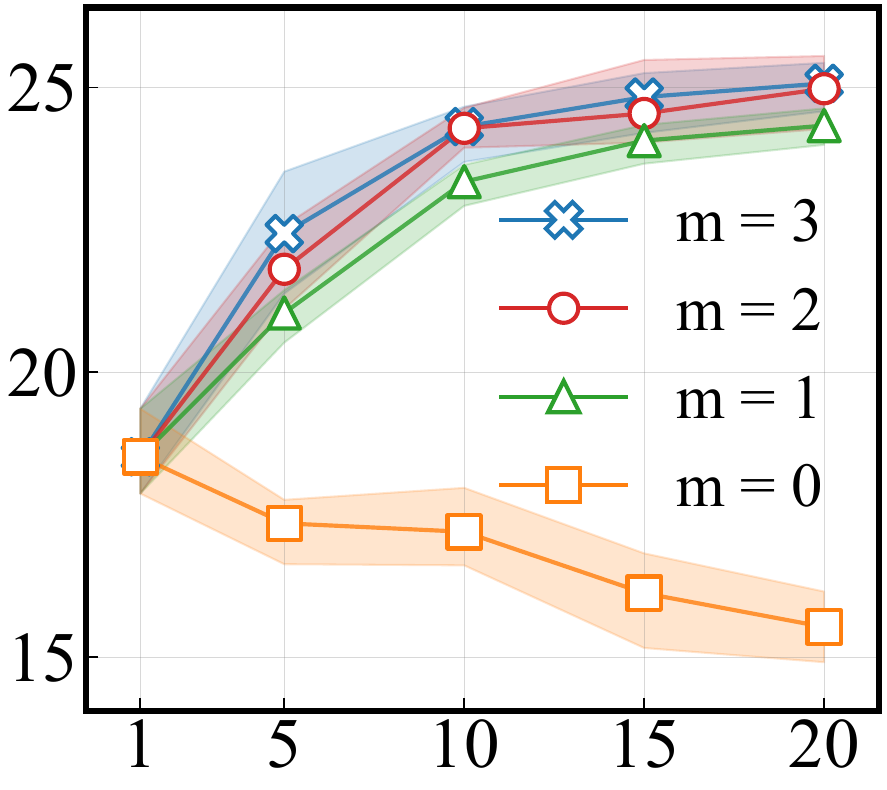}{0.94} & \entry{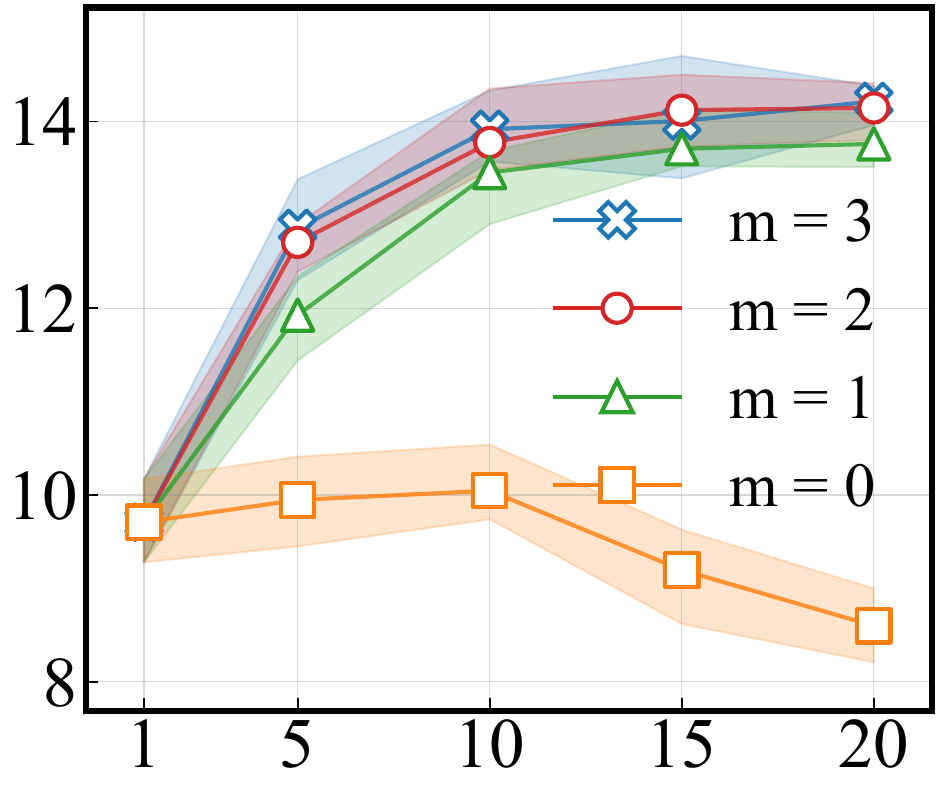}{1.0} \\
    & \multicolumn{2}{c}{\xlabel} \\
  \end{tabular}
\caption{
Effect of local-guidance refinement under different window sizes.
$m{=}0$ does not refine the inherited guidance and updates it only when the guidance runs out.
$m{=}1$ updates the inherited guidance once per step, and so force.
}
\label{fig:refinement}
\end{figure}
}

\paragraph{Anytime Refinement with Guidance.}
The preceding results show that LLLG is sufficiently fast for real-time lifelong operation.
This leaves room to apply anytime planning within the per-step computation budget.
LLLG admits two such refinement options, LaCAM$^\ast$ and LNS; here we focus on LaCAM$^\ast$, and defer the LNS results to the appendix.

After finding an initial $w_{\Pi}$-step plan, LaCAM$^\ast$ continues the search within the same finite horizon and seeks to improve it.
As shown in \cref{fig:anytime}, the windowed version of lifelong LaCAM exhibits a small improvement with anytime refinement, whereas the same refinement degrades LLLG's performance.
This suggests a discrepancy between the high-level cost function used by LaCAM$^*$ (i.e., $f$-value improvement on the $w_{\Pi}$-step window) and the lifelong performance metric of interest (throughput).
A similar tendency is observed for LNS: its benefit is map-dependent, improving performance on some smaller-scale maps but diminishing or even becoming negative in dense settings.

{
\setlength{\tabcolsep}{2pt}
\newcommand{\realtimefig}[1]{fig/raw/anytime/runtime_throughput_#1}

% method label positions (x,y) in normalized image coordinates
\newcommand{\methodposSmallLaCAM}{{0.09,0.83}}
\newcommand{\methodposSmallLG}{{0.51,0.65}}
\newcommand{\methodposSmallLGRefined}{{0.46,0.65}}

\newcommand{\methodposLargeLaCAM}{{0.13,0.75}}
\newcommand{\methodposLargeLG}{{0.58,0.65}}
\newcommand{\methodposLargeLGRefined}{{0.55,0.54}}

\newcommand{\methodlabel}[5]{% #1: panel key, #2: method key, #3: color, #4: label, #5: anchor
  \node[anchor=#5] at (\csname methodpos#1#2\endcsname) {\scriptsize \textcolor[HTML]{#3}{\shortstack[c]{#4}}};
}
\newcommand{\methodlabels}[1]{% #1: panel key
  \methodlabel{#1}{LaCAM}{2ca02c}{LaCAM}{west}
  \methodlabel{#1}{LG}{1f77b4}{LG}{west}
  \methodlabel{#1}{LGRefined}{d62728}{LG-Refined}{west}
}

\newcommand{\ylabel}{\rotatebox{90}{\small \hspace{1.5em}{ throughput $\rightarrow$}}}
\newcommand{\xlabel}{\small $\leftarrow$ runtime [\SI{}{\second}]}

\newcommand{\agentlegendSmall}{% [100, 200, 300, 400]
  \begin{scope}[overlay,shift={(1.03,-0.09)},scale=0.85,transform shape]
    \draw[] (0, 0) --
    node[pos=0]{\tiny $\times$}
    node[pos=0.333]{\tiny $\bullet$}
    node[pos=0.666]{\tiny $\blacktriangle$}
    node[pos=1.0]{\tiny $\blacksquare$}
    node[right,pos=0]{\tiny 100}
    node[right,pos=0.333]{\tiny 200}
    node[right,pos=0.666]{\tiny 300}
    node[right,pos=1]{\tiny 400}
    node[above right=0.2,pos=1]{\tiny agents}
    (0.06, 0.24);
  \end{scope}%
}
\newcommand{\agentlegendLarge}{% [200, 400, 600, 800, 1000]
  % place outside bottom-right of each panel
  \begin{scope}[overlay,shift={(1.03,-0.09)},scale=0.85,transform shape]
    \draw[] (0, 0) --
    node[pos=0]{\tiny $\times$}
    node[pos=0.25]{\tiny $\bullet$}
    node[pos=0.5]{\tiny $\blacktriangle$}
    node[pos=0.75]{\tiny $\blacksquare$}
    node[pos=1.0]{\tiny $\bigstar$}
    node[right,pos=0]{\tiny 200}
    node[right,pos=0.25]{\tiny 400}
    node[right,pos=0.5]{\tiny 600}
    node[right,pos=0.75]{\tiny 800}
    node[right,pos=1]{\tiny 1000}
    node[above right=0.2,pos=1]{\tiny agents}
    (0.06, 0.24);
  \end{scope}%
}

\newcommand{\entry}[2]{% #1: instance key, #2: panel key
  \begin{minipage}[t]{0.45\linewidth}
    \centering
    \begin{tikzpicture}
      \node[anchor=south west,inner sep=0] (img) at (0, 0) {%
        \includegraphics[width=0.78\linewidth]{\realtimefig{#1}}%
      };
      \begin{scope}[x={(img.south east)},y={(img.north west)}]
        % draw labels per-figure
        \methodlabel{#2}{LaCAM}{008000}{Windowed\\LaCAM}{west}
        \methodlabel{#2}{LG}{0000FF}{LLLG}{west}
        \node[overlay,anchor=north west] at (0.18,0.0) {\xlabel};
        \csname agentlegend#2\endcsname
      \end{scope}
    \end{tikzpicture}
    \vspace{1.0em}
  \end{minipage}%
}
\begin{figure}[t!]
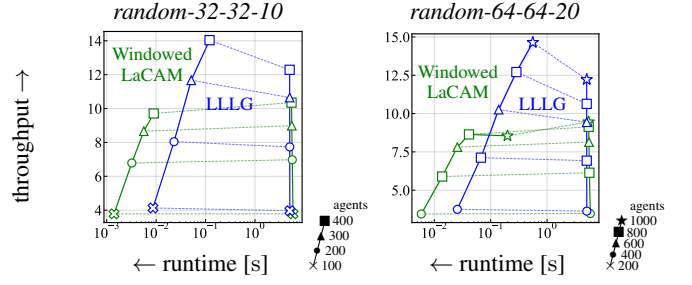

  \centering
  \begin{tabular}{@{}c@{\hspace{0.8em}}cc@{}}
    & \small \mapname{random-32-32-10} & \small \mapname{random-64-64-20} \\
    \ylabel &
    \entry{random-32-32-10}{Small} &
    \entry{random-64-64-20}{Large} \\
  \end{tabular}
  \caption{
    Effect of anytime improvement on the algorithms.
    The planning runtime budget is $5$ \si{\second}/step in both maps.
    Dashed lines connect each algorithm to its anytime variant.
  }
  \label{fig:anytime}
\end{figure}
}

\section{Discussion}
This study shows that local guidance improves solution quality in lifelong MAPF, boosting throughput under real-time budgets.
Integrated into a receding-horizon loop, our LLLG remains fast, scales to large maps, and mitigates dense bottlenecks.
A key design choice is how to leverage the previous planning effort.
Rather than warm-starting the finite-horizon solution itself, we warm-start the local guidance from the previous-step solution, whose near-term collision-free forecast is more informative than inherited guidance alone.

Empirically, increasing the planning horizon $w_{\Pi}$ and the guidance window size $w_{\Phi}$ yields diminishing returns, and throughput improvement appears to saturate even with larger windows.
A plausible explanation is that, in LMAPF, future goal assignments after task completion are unknown; thus, longer-horizon planning increasingly optimizes predictions that become irrelevant once agents reach their current goals and are reassigned.
Addressing this limitation may require information about future tasks, such as future goal assignments or a predictable goal distribution.
Exploring such extensions is beyond the scope of this paper, but it remains a key direction for the future.
It is also important to explore cost evaluation for anytime refinement whose finite-horizon improvements better align with lifelong throughput.
\section*{Acknowledgments}
This research was supported by a gift from Murata Machinery, Ltd.
\bibliography{sty/ref-macro,ref}

\appendix

\section{Appendix}
\suppressfloats[t]

\paragraph{LNS is not reliably effective.}
Improving solution quality in lifelong MAPF via LNS is an important question. However, in our lifelong design, applying LNS at each step can only refine the predicted finite-horizon plan, while the executed trajectory is committed online and cannot be retroactively revised. Therefore, it is not obvious that per-step LNS refinement should translate directly into better lifelong performance.
\Cref{fig:lns2} suggests a nuanced conclusion. On smaller-scale maps, LNS tends to improve performance, indicating that refining the short-horizon plan can indeed reduce inefficiencies that accumulate over time. 
In contrast, this benefit is not consistent across all maps. 
In dense settings with many agents, the marginal gain from LNS diminishes and can even become negative, likely because additional search struggles to meaningfully restructure congestion-heavy interactions within a tight step budget. 
{
\setlength{\tabcolsep}{2pt}
\newcommand{\realtimefig}[1]{fig/raw/lns/runtime_throughput_#1}

% method label positions (x,y) in normalized image coordinates
\newcommand{\methodposSmallLaCAM}{{0.12,0.75}}
\newcommand{\methodposSmallLG}{{0.64,0.81}}
\newcommand{\methodposSmallLGRefined}{{0.58,0.62}}

\newcommand{\methodposLargeLaCAM}{{0.17,0.65}}
\newcommand{\methodposLargeLG}{{0.65,0.71}}
\newcommand{\methodposLargeLGRefined}{{0.59,0.50}}

\newcommand{\methodlabel}[5]{% #1: panel key, #2: method key, #3: color, #4: label, #5: anchor
  \node[anchor=#5] at (\csname methodpos#1#2\endcsname) {\scriptsize \textcolor[HTML]{#3}{\shortstack[c]{#4}}};
}
\newcommand{\methodlabels}[1]{% #1: panel key
  \methodlabel{#1}{LaCAM}{2ca02c}{LaCAM}{west}
  \methodlabel{#1}{LG}{1f77b4}{LG}{west}
  \methodlabel{#1}{LGRefined}{d62728}{LG-Refined}{west}
}

\newcommand{\ylabel}{\rotatebox{90}{\small \hspace{1.5em}{ throughput $\rightarrow$}}}
\newcommand{\xlabel}{\small  $\leftarrow$ runtime [\SI{}{\second}]}

\newcommand{\agentlegendSmall}{% [100, 200, 300, 400]
  \begin{scope}[overlay,shift={(1.03,-0.09)},scale=0.85,transform shape]
    \draw[] (0, 0) --
    node[pos=0]{\tiny $\times$}
    node[pos=0.333]{\tiny $\bullet$}
    node[pos=0.666]{\tiny $\blacktriangle$}
    node[pos=1.0]{\tiny $\blacksquare$}
    node[right,pos=0]{\tiny 100}
    node[right,pos=0.333]{\tiny 200}
    node[right,pos=0.666]{\tiny 300}
    node[right,pos=1]{\tiny 400}
    node[above right=0.2,pos=1]{\tiny agents}
    (0.06, 0.24);
  \end{scope}%
}
\newcommand{\agentlegendLarge}{% [200, 400, 600, 800, 1000]
  % place outside bottom-right of each panel
  \begin{scope}[overlay,shift={(1.03,-0.09)},scale=0.85,transform shape]
    \draw[] (0, 0) --
    node[pos=0]{\tiny $\times$}
    node[pos=0.25]{\tiny $\bullet$}
    node[pos=0.5]{\tiny $\blacktriangle$}
    node[pos=0.75]{\tiny $\blacksquare$}
    node[pos=1.0]{\tiny $\bigstar$}
    node[right,pos=0]{\tiny 200}
    node[right,pos=0.25]{\tiny 400}
    node[right,pos=0.5]{\tiny 600}
    node[right,pos=0.75]{\tiny 800}
    node[right,pos=1]{\tiny 1000}
    node[above right=0.2,pos=1]{\tiny agents}
    (0.06, 0.24);
  \end{scope}%
}

\newcommand{\entry}[2]{% #1: instance key, #2: panel key
  \begin{minipage}[t]{0.45\linewidth}
    \centering
    \begin{tikzpicture}
      \node[anchor=south west,inner sep=0] (img) at (0, 0) {%
        \includegraphics[width=0.78\linewidth]{\realtimefig{#1}}%
      };
      \begin{scope}[x={(img.south east)},y={(img.north west)}]
        % draw labels per-figure
        \methodlabel{#2}{LaCAM}{008000}{LLLG\\(w/o $\boldsymbol{\Phi})$}{west}
        \methodlabel{#2}{LG}{0000FF}{LLLG}{west}
        \node[overlay,anchor=north west] at (0.18,0.0) {\xlabel};
        \csname agentlegend#2\endcsname
      \end{scope}
    \end{tikzpicture}
    \vspace{1.0em}
  \end{minipage}%
}
\begin{figure}[h!]
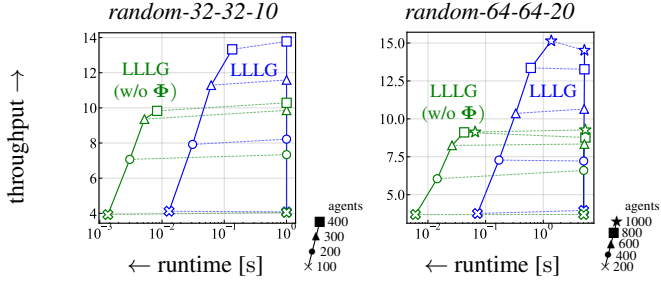

  \centering
  \begin{tabular}{@{}c@{\hspace{0.8em}}cc@{}}
    & \small \mapname{random-32-32-10} & \small \mapname{random-64-64-20} \\
    \ylabel &
    \entry{random-32-32-10}{Small} &
    \entry{random-64-64-20}{Large} \\
  \end{tabular}
  \caption{
    % LNS.
    Effect of LNS improvement for different lifelong algorithms.
    The planning runtime budget including LNS is $1$ s/step (left) and $5$ s/step (right).
    Dashed lines connect each algorithm to its LNS-improved variant.
    LLLG(w/o $\boldsymbol{\Phi}$) corresponds to LaCAM.
  }
  \label{fig:lns2}
\end{figure}
}

\paragraph{Hindrance is a practical option.}
We also examine the effect of integrating hindrance.
Especially in highly dense settings such as shown in \Cref{fig:hindrance}, combining hindrance with LLLG improves throughput with negligible computation cost.
Thus, hindrance is a lightweight enhancement that is generally worth enabling in practice.

{
\setlength{\tabcolsep}{2pt}
\newcommand{\windowfig}[2]{fig/raw/hindrance/hindrance_#1_#2agents}

% window label positions (x,y) in normalized image coordinates (per panel)
\newcommand{\labelposWindowSmallA}{{0.16,0.23}}
\newcommand{\labelposWindowSmallB}{{0.3,0.6}}
\newcommand{\labelposWindowSmallC}{{0.38,0.63}}
\newcommand{\labelposWindowSmallD}{{0.22,0.9}}

\newcommand{\labelposWindowLargeA}{{0.25,0.23}}
\newcommand{\labelposWindowLargeB}{{0.21,0.48}}
\newcommand{\labelposWindowLargeC}{{0.38,0.68}}
\newcommand{\labelposWindowLargeD}{{0.22,0.94}}

\newcommand{\labelposWindowMediumA}{{0.16,0.23}}
\newcommand{\labelposWindowMediumB}{{0.3,0.68}}
\newcommand{\labelposWindowMediumC}{{0.38,0.64}}
\newcommand{\labelposWindowMediumD}{{0.18,0.9}}

\newcommand{\windowlabel}[5]{% #1: panel key, #2: label key, #3: color, #4: label, #5: anchor
  \node[anchor=#5] at (\csname labelposWindow#1#2\endcsname) {\scriptsize \textcolor[HTML]{#3}{\textbf{#4}}};
}
\newcommand{\windowlabels}[1]{% #1: panel key
  % \windowlabel{#1}{A}{ff7f0e}{$w_{\Phi}$=3}{west}
  % \windowlabel{#1}{B}{2ca02c}{$w_{\Phi}$=8}{west}
  \windowlabel{#1}{C}{d62728}{w/o hindrance}{west}
  \windowlabel{#1}{D}{1f77b4}{w/ hindrance}{west}
}
\newcommand{\agentlegendLarge}{% [200, 400, 600, 800, 1000]
  \begin{scope}[shift={(0.55,0.15)}]
    \draw[] (0, 0) --
    node[pos=0]{\tiny $\times$}
    node[pos=0.25]{\tiny $\bullet$}
    node[pos=0.5]{\tiny $\blacktriangle$}
    node[pos=0.75]{\tiny $\blacksquare$}
    node[pos=1.0]{\tiny $\bigstar$}
    node[right,pos=0]{\tiny 3}
    node[right,pos=0.25]{\tiny 5}
    node[right,pos=0.5]{\tiny 8}
    node[right,pos=0.75]{\tiny 15}
    node[right,pos=1]{\tiny $w_{\Phi}$=20}
    %
    % node[above right=0.2,pos=1]{\tiny $w_{\Pi}$}
    (0.06, 0.24);
  \end{scope}%
}
\newcommand{\agentlegendSmall}{%
  \agentlegendLarge%
}
\newcommand{\agentlegendMedium}{%
  \agentlegendLarge%
}
\newcommand{\entry}[4]{% #1: map, #2: agents, #3: panel key, #4: image width
  \begin{minipage}[b]{0.35\linewidth}
    \centering
    \parbox[b][1.8em][c]{\linewidth}{\centering \small \mapname{#1}}\par
    \vspace{0.2em}
    \begin{tikzpicture}
      \node[anchor=south west,inner sep=0] (img) at (0, 0) {%
        \includegraphics[width=#4\linewidth]{\windowfig{#1}{#2}}%
      };
      \begin{scope}[x={(img.south east)},y={(img.north west)}]
        \windowlabels{#3}
        \csname agentlegend#3\endcsname
      \end{scope}
    \end{tikzpicture}
  \end{minipage}%
}
\newcommand{\ylabel}{\rotatebox{90}{\small \hspace{4.0em}{ throughput $\rightarrow$}}}
\newcommand{\xlabel}{\small  $\leftarrow$ runtime [\SI{}{\second}]}
\begin{figure}[h!]
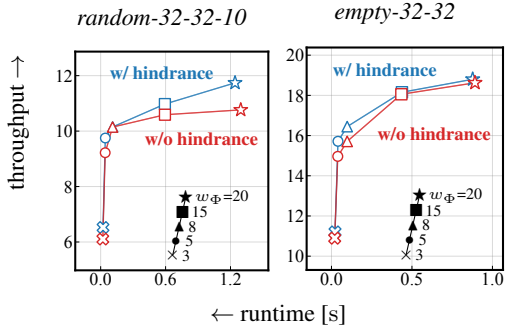

  \centering
  \begin{tabular}{@{}c@{\hspace{0.8em}}cc@{}}
    \ylabel &
    \entry{random-32-32-10}{1000}{Small}{0.98} &
    \entry{empty-32-32}{1000}{Medium}{1.0} \\
    % &
    % \entry{random-64-64-20}{200}{Medium}{0.95} &
    % \entry{random-64-64-20}{1000}{Large}{1.0} \\
    & \multicolumn{2}{c}{\xlabel} \\
  \end{tabular}
\caption{
Sensitivity of throughput and runtime to the hindrance and the guidance window size $w_{\Phi}$.
Results is evaluated on \mapname{random-32-32-10} and \mapname{empty-32-32} with $1,000$ agents.
}
\label{fig:hindrance}
\end{figure}
}

\end{document}